\documentclass[aps,prx,10pt,onecolumn,showpacs,showkeys,amsmath,amssymb,superscriptaddress]{revtex4-2}
\usepackage{graphicx}
\usepackage{dcolumn}
\usepackage{bm}
\usepackage[version=3]{mhchem}
\usepackage{hyperref}
\hypersetup{colorlinks,citecolor=blue,filecolor=blue,linkcolor=blue,urlcolor=blue}
\usepackage{booktabs}
\usepackage[utf8]{inputenc}
\usepackage{xcolor}
\DeclareUnicodeCharacter{2212}{-}
\DeclareUnicodeCharacter{2009}{\,}

\usepackage{caption}
\DeclareCaptionLabelFormat{suppfig}{\textbf{Supplementary Figure #2}}
\DeclareCaptionLabelFormat{supptab}{\textbf{Supplementary Table #2}}
\DeclareCaptionLabelSeparator{pipe}{\textbf{ | }}

\begin{document}

\title{Localized Excitons and Exciton–Phonon Coupling in Antiferromagnetic AgCrP$_2$S$_6$}
\author{Jessica McDivitt}
\affiliation{University of Colorado Boulder, Boulder, CO 80309, USA}
\author{Dimitar Pashov}
\affiliation{Theory and Simulation of Condensed Matter, King's College London, The Strand, London WC2R 2LS, United Kingdom}
\author{Mark van Schilfgaarde}
\affiliation{National Laboratory of the Rockies, Golden, CO 80401, USA}
\author{Justin C. Johnson}
\affiliation{National Laboratory of the Rockies, Golden, CO 80401, USA}
\author{Jeffrey L. Blackburn}
\affiliation{National Laboratory of the Rockies, Golden, CO 80401, USA}
\author{Anna A. Berseneva}
\affiliation{National Laboratory of the Rockies, Golden, CO 80401, USA}
\author{Swagata Acharya}
\email{swagata.acharya@nlr.gov}
\affiliation{National Laboratory of the Rockies, Golden, CO 80401, USA}
\begin{abstract}
AgCrP$_2$S$_6$ combines a low-symmetry thiophosphate framework with an antiferromagnetic Cr sublattice and nonmagnetic Ag sites, providing a setting in which covalency and magnetic localization compete in the low-energy optical response. Single-crystal x-ray diffraction, Raman spectroscopy, lattice-dynamical calculations, and self-consistent vertex-corrected Feynman diagrammatic many-body approaches were used to determine the structural, electronic, and excitonic properties of bulk AgCrP$_2$S$_6$. The material remained monoclinic between 100 and 300 K, with additional Ag-site disorder resolved at low temperatures, and the optimized structure was dynamically stable. The many-body calculations yield a reduced quasiparticle gap relative to Cr trihalides, but the lowest excitons remain predominantly local, and the calculated low-energy ($\sim$1.4 eV) excitation cluster is best described as a weakly bright, strongly anisotropic Frenkel exciton with a dominant onsite $d$--$d$ character and substantial ligand-assisted $d$--$p$ admixture. Symmetry analysis in the $C_{2h}$ setting identifies this state as predominantly $B_u$ and constrains its leading exciton-phonon coupling channels to $A_g$ diagonal renormalization and $B_g$-mediated bright-dark mixing. These results place AgCrP$_2$S$_6$ in a localized excitonic regime in which the stronger $p$–d hybridization, relative to the more ionic Cr trihalides, narrows the quasiparticle gap, while the antiferromagnetic exchange and Ag-site dilution disfavor the intersite coherence associated with strongly delocalized low-energy excitons. Benchmarked against Cr-, Ni-, and Mn-based thiophosphates and chromium trihalides of differing local symmetry and crystal field, this behavior identifies cation substitution and ligand-ion engineering as a general, transferable handle for tuning the degree of exciton localization across this broader materials family, with direct consequences for how these excitons couple to bosons and for their potential utility in spintronic applications.
\end{abstract}

\maketitle

\section*{Introduction}

Atomically thin crystals and van der Waals heterostructures provide a tunable platform in which magnetic exchange, structural symmetry, and reduced dimensionality can be engineered together within a single materials class~\cite{novoselov2004electric,geim2013van,park2026vdwmagnets}. The discovery of intrinsic magnetism in monolayer CrI$_3$ and Cr$_2$Ge$_2$Te$_6$ established a broader platform for magneto-optical spectroscopy in van der Waals compounds~\cite{Gong2017Discovery,Huang2017Layer}. Recent reviews emphasized that excitons in van der Waals magnets must be understood in terms of exchange, localization, and exciton binding on comparable energy scales~\cite{park2026vdwmagnets,adak2026vdwreview}. In chromium trihalides, ligand-field excitations, dark-bright exciton structures, and magneto-optical responses depend sensitively on the local coordination and magnetic order~\cite{Burch2018Magnetooptical,Seyler2018LigandField,molina2020magneto,wu2019physical,PhysRevMaterials.6.014008,acharya2021electronic,acharya2022real}. Resonant inelastic x-ray scattering has reinforced the same message from a complementary angle by directly resolving local $d$--$d$ excitations and their magneto-structural dressing in CrI$_3$ and related systems~\cite{ghosh2023magnetic,RevModPhys.83.705}. More recently, direct coupling between excitons, exchange splitting, and magnetic collective effects have been resolved in anisotropic and antiferromagnetic van der Waals magnets, including CrSBr~\cite{Klein_2023,Wilson_2021,watson2024giant,ruta2023hyperbolic,shao2024exciton,datta2024magnon}. Metal thio- and selenophosphates extend this landscape by combining low-symmetry, mixed covalent-ionic bonding, and chemically tunable magnetic sublattices within a common layered framework~\cite{Wang2022Review}. Within this family, members such as NiPS$_3$, MnPS$_3$, and CoPS$_3$ have established connections between the local orbital character, spin--lattice coupling, and spin-entangled optical excitations~\cite{kang2020coherent,he2024magnetically, kim2021charge,scheie2023spin,lane2020spin, khusyainov2023ultrafast,jana2025deconstruction,na2026engineering}.

Microscopically, the exciton landscape in magnets with partially filled $d$ shells is governed by a combination of local ligand-field excitations comprising onsite $d\text{--}d$, intersite $d\text{--}d$, and hybrid $d\text{--}p$~\cite{wu2019physical,adak2026vdwreview} transitions. How bright or dark a given exciton is depends on its orbital and spin character, as shown for CrX$_3$ by Wu \textit{et al.}~\cite{wu2019physical,PhysRevMaterials.6.014008}, and on the parity of the two-particle wavefunction on the magnetic sublattices~\cite{acharya2026brightdarkexcitonscrsbr}. As a representative layered antiferromagnetic semiconductor containing a nonmagnetic metal cation, $\mathrm{AgCrP_2S_6}$~\cite{park2024anisotropic,selter2021crystal,mutka1993one,PhysRevB,rao2025chiral} provides an ideal model system to explore this behavior. In $\mathrm{AgCrP_2S_6}$, replacing half of the magnetic Cr network with Ag alters both the local site symmetry and the intersite dipole interference, fundamental parameters that reshape these bright and dark transition manifolds.   In such systems, two competing effects dictate the interplay between excitons and the underlying magnetic background: while the sulfur-phosphorus framework and reduced symmetry favor extended hybridized orbitals, the Cr network is antiferromagnetic, which restricts carrier hopping relative to a ferromagnetic chain, and filled-shell Ag$^+$ blocks Cr$\to$Ag $d$--$d$ paths, as discussed below. $\mathrm{AgCrP_2S_6}$ is therefore a useful counterpoint to other Cr-based layered magnets in which low-energy excitons retain a significant amount of intersite character \cite{acharya2021electronic,acharya2022real}. It also connects to the broader question of whether the weak optical response in correlated insulators originates from spin-forbidden transitions or from non-spin-flip local excitons whose oscillator strength is suppressed by symmetry and reduced intersite coherence \cite{jana2025deconstruction,acharyaTheoryColorsStrongly2023}. The central question, then, for antiferromagnetic semiconductors such as $\mathrm{AgCrP_2S_6}$, is how much local Frenkel exciton character survives when covalency increases, symmetry is lowered, and the magnetic network limits coherent intersite transport.

We combine structural characterization and temperature-dependent spectroscopy with self-consistent many-body calculations~\cite{Cunningham2023,qsgw,questaal_paper}. Here QS$GW$ denotes quasiparticle self-consistent $GW$, and QS$G\hat{W}$ is its electron--hole vertex-corrected self-consistent extension in which electron--hole correlations are included through ladder vertex corrections obtained by solving the Bethe--Salpeter equation (BSE). Earlier work has reported the crystal chemistry, anisotropic optics, and phonons of AgCrP$_2$S$_6$~\cite{park2024anisotropic,selter2021crystal,Auti_Antipolar,volochanskyi_indirect_2026}. What remains open is how AFM order and Ag dilution set the localization of the lowest exciton, how that exciton contrasts with those reported in other 2D magnets, and which Raman-active phonon symmetries dress that state. We address those points in the Results through the QS$G\hat{W}$ quasiparticle spectrum and BSE exciton decomposition, the $C_{2h}$ exciton--phonon selection rules, and the photoluminescence analysis in Figs.~\ref{fig:crystal-electronic}--\ref{fig:pl} and Tables~\ref{tab:crystdata}, \ref{tab:atomcoords}, \ref{tab:exciton-phonon-symmetry}, and \ref{tab:raman-pl-modes}.

\section*{Results}

\subsection*{Crystal and Electronic Structure}

\begin{figure*}[t]
\centering
\includegraphics[width=0.99\linewidth]{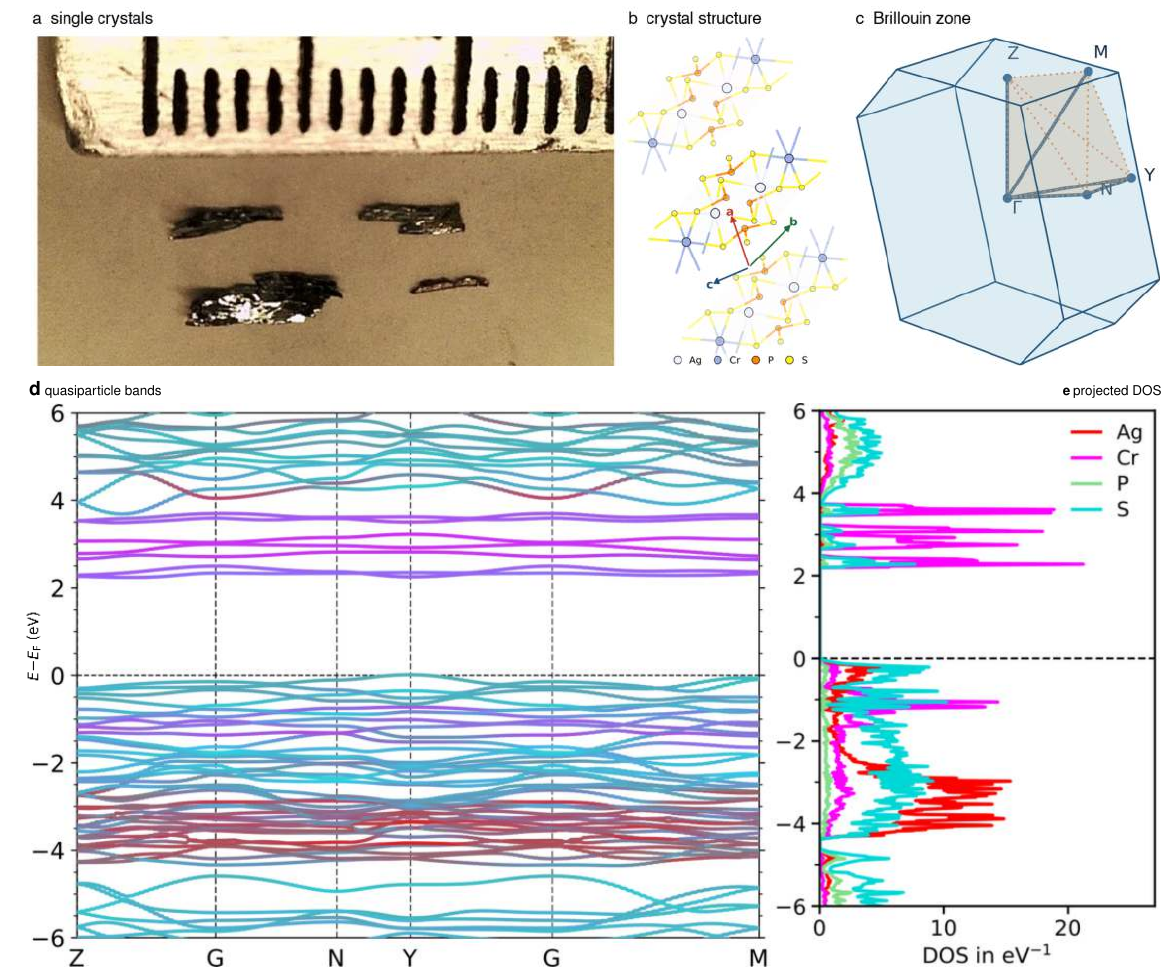}
\caption{Crystal and electronic structure of AgCrP$_2$S$_6$. (a) Representative crystals used for diffraction (full growth images in the Supplementary Information). (b) Crystal structure in an oblique projection, with Ag/Cr/P/S colors and crystallographic $a$, $b$, $c$ axes indicated; neighbouring layers are faded for clarity but drawn at comparable marker size. (c) First Brillouin zone and high-symmetry path. (d) Quasiparticle band structure, with orbital/atom colouring matching the projected-DOS legend in (e); the shared vertical axis is energy relative to the Fermi level, $E-E_{\mathrm{F}}$ (eV). (e) Atom-projected electronic density of states. The structure colours in (b) are reused in the real-space exciton visualizations below.}
\label{fig:crystal-electronic}
\end{figure*}

Single crystals of $\mathrm{AgCrP_2S_6}$ form as elongated needles with lengths of up to 10~mm [Fig.~\ref{fig:crystal-electronic}(a)]. Single-crystal X-ray diffraction confirmed that the material remained monoclinic ($P2/c$) from 300~K down to 100~K, where explicit Ag-site splitting accounts for low-temperature local disorder (Tables~\ref{tab:crystdata} and \ref{tab:atomcoords}). The crystal structure, shown in Fig.~\ref{fig:crystal-electronic}(b), features a layered thiophosphate framework with an ordered Ag/Cr stripe motif in a monoclinic setting. The corresponding first Brillouin zone and the high-symmetry $\mathbf{k}$-path used for energy band dispersion are depicted in Fig.~\ref{fig:crystal-electronic}(c).  All quasiparticle and BSE calculations use this refined monoclinic cell and the experimental antiferromagnetic Cr order (Methods). The Ag/Cr stripe motif is part of the measured crystal structure. To establish how this low-symmetry layered framework shapes the electronic spectrum, we evaluated the quasiparticle band structure across increasing levels of electronic correlation within the many-body perturbation theory.

\begin{table*}[t]
\caption{Crystallographic data for Ag$_{0.5}$Cr$_{0.5}$PS$_3$.}
\label{tab:crystdata}
\begin{tabular*}{\textwidth}{@{\extracolsep{\fill}}lcc@{}}
\hline
Parameter & Ag$_{0.5}$Cr$_{0.5}$PS$_3$ (100 K) & Ag$_{0.5}$Cr$_{0.5}$PS$_3$ (300 K) \\
\hline
Empirical formula & Ag$_{0.5}$Cr$_{0.5}$PS$_3$ & Ag$_{0.5}$Cr$_{0.5}$PS$_3$ \\
Formula weight & 207.09 & 207.09 \\
Temperature (K) & 100 & 300 \\
Crystal system & Monoclinic & Monoclinic \\
Space group & $P2/c$ & $P2/c$ \\
$a$ (\AA) & 6.7188(4) & 6.7475(4) \\
$b$ (\AA) & 10.6025(6) & 10.6327(5) \\
$c$ (\AA) & 5.8642(3) & 5.8858(3) \\
$\alpha$ (deg) & 90 & 90 \\
$\beta$ (deg) & 105.902(2) & 106.019(2) \\
$\gamma$ (deg) & 90 & 90 \\
Volume (\AA$^3$) & 401.76(4) & 405.87(4) \\
$Z$ & 4 & 4 \\
$\rho_{\rm calc}$ (g/cm$^3$) & 3.424 & 3.389 \\
$\mu$ (mm$^{-1}$) & 33.009 & 32.674 \\
$F(000)$ & 394 & 394 \\
Crystal size (mm$^3$) & $0.05 \times 0.03 \times 0.005$ & $0.05 \times 0.03 \times 0.005$ \\
$2\theta$ range (deg) & 13.964 to 113.914 & 13.916 to 113.916 \\
Index ranges & $-8 \le h \le 8$, $-11 \le k \le 13$, $-7 \le l \le 7$ & $-8 \le h \le 8$, $-12 \le k \le 13$, $-7 \le l \le 7$ \\
Reflections collected & 9528 & 12009 \\
Data/restraints/parameters & 809/1/58 & 827/0/47 \\
Goodness-of-fit on $F^2$ & 1.081 & 1.055 \\
$R_1/wR_2$ [$I \ge 2\sigma(I)$] (\%) & 3.50/9.84 & 3.20/9.20 \\
$R_1/wR_2$ (all data) (\%) & 3.58/9.97 & 3.44/9.43 \\
Largest diff. peak/hole (e \AA$^{-3}$) & 1.01/--0.81 & 1.08/--0.73 \\
$R_{\rm int}$ (\%) & 4.56 & 4.42 \\
\hline
\end{tabular*}
\end{table*}

\begin{table*}[t]
\caption{Fractional atomic coordinates ($\times 10^4$) and equivalent isotropic displacement parameters (\AA$^2 \times 10^3$) for Ag$_{0.5}$Cr$_{0.5}$PS$_3$. $U_{\rm eq}$ is defined as one third of the trace of the orthogonalized $U_{IJ}$ tensor.}
\label{tab:atomcoords}
\begin{tabular*}{\textwidth}{@{\extracolsep{\fill}}llccccc@{}}
\hline
Temperature & Atom & $x$ & $y$ & $z$ & $U_{\rm eq}$ & Occ. \\
\hline
100 K & Ag1 & 5000 & 4375(16) & 7500 & 15(4) & 0.380(3) \\
100 K & Ag2 & 5120(50) & 4366(10) & 7853(14) & 8(2) & 0.1116(17) \\
100 K & Ag3 & 4750(20) & 4365(15) & 7470(50) & 8(2) & 0.1734(17) \\
100 K & Ag4 & 5780(40) & 4380(20) & 7810(40) & 8(2) & 0.0248(17) \\
100 K & Cr1 & 5000 & 9222.4(10) & 2500 & 11.3(3) & 1.0 \\
100 K & P1 & 6704.3(18) & 7541.0(10) & 7982.5(19) & 11.9(3) & 1.0 \\
100 K & S1 & 7399.5(18) & 7709.7(11) & 4836.0(19) & 13.5(3) & 1.0 \\
100 K & S2 & 7868.7(17) & 6041.0(12) & 9832.5(18) & 14.5(3) & 1.0 \\
100 K & S3 & 7302.9(17) & 9231.3(10) & 9923.9(19) & 12.5(3) & 1.0 \\
\hline
300 K & Ag1 & 5000 & 4366.3(5) & 7500 & 49.0(2) & 1.0 \\
300 K & Cr1 & 5000 & 9214.9(8) & 2500 & 20.3(3) & 1.0 \\
300 K & P1 & 6694.8(15) & 7546.2(9) & 7980.9(16) & 21.0(3) & 1.0 \\
300 K & S1 & 7394.4(15) & 7709.7(10) & 4851.4(16) & 24.6(3) & 1.0 \\
300 K & S2 & 7860.9(16) & 6052.0(11) & 9827.1(16) & 29.0(3) & 1.0 \\
300 K & S3 & 7286.8(14) & 9234.7(9) & 9924.8(15) & 21.8(3) & 1.0 \\
\hline
\end{tabular*}
\end{table*}

 Within the present calculation, the band gap evolves from 1.16\,eV in LDA to 2.42\,eV in $\mathrm{QS}GW$ and 2.19\,eV in $\mathrm{QS}G\hat{W}$ (see Method section). For all discussions, the $\mathrm{QS}G\hat{W}$ electronic structures and excitonic spectra are used in the current work. The electronic spectrum is detailed by the quasiparticle band structure in Fig.~\ref{fig:crystal-electronic}(d) and the atom-projected density of states in Fig.~\ref{fig:crystal-electronic}(e). In the present QS$G\hat{W}$ description, the top of the valence bands is formed from strongly hybridized Cr-$d$/S-$p$ states (with the ligand S-$p$ states dominating the valence edges), while the lowest conduction states retain substantial Cr and ligand character rather than becoming purely ionic. This hybridized edge structure is crucial for understanding why AgCrP$_2$S$_6$ does not reduce to an isolated atomic $d$--$d$ excitation problem (discussed later). The occupied Cr-$t_{2g}$ states form a manifold with a center of mass positioned roughly 1~eV below the valence band maximum, while the Ag-derived states reside deeper in the valence manifold, with their center of mass centered around 3.5~eV below $E_F$. In that sense, the valence band electronic structure of AgCrP$_2$S$_6$ has strong similarities with NiPS$_{3}$ and MnPS$_{3}$~\cite{jana2025deconstruction} and their alloys~\cite{na2026engineering}. 

 The overall band gap magnitude remains smaller than that in the Cr trihalides studied with the same many-body framework~\cite{acharya2021electronic}. The sulfur-phosphorus framework and low (monoclinic) symmetry broaden the hybridized molecular-orbital network and reduce the gap relative to the more ionic CrX$_3$ family. AFM order on the Cr network acts against that covalency-driven delocalization, so the low-energy carriers remain more local than the reduced gap alone would suggest; the excitonic consequences are developed below. The band edges are also comparatively flat along the $Z$-$\Gamma$-$N$-$Y$-$\Gamma$-$M$ path, consistent with a near-direct or weakly indirect edge~\cite{volochanskyi_indirect_2026} landscape rather than a single sharply isolated direct-gap extremum. AgCrP$_2$S$_6$ therefore occupies a regime where structural delocalization and magnetic localization compete on comparable footing, and where the lowest optical exciton is selected from a shallow band-edge manifold rather than from a strongly allowed direct transition.

\subsection*{Exciton Character and Localization}

\begin{figure*}[t]
\centering
\includegraphics[width=0.98\linewidth]{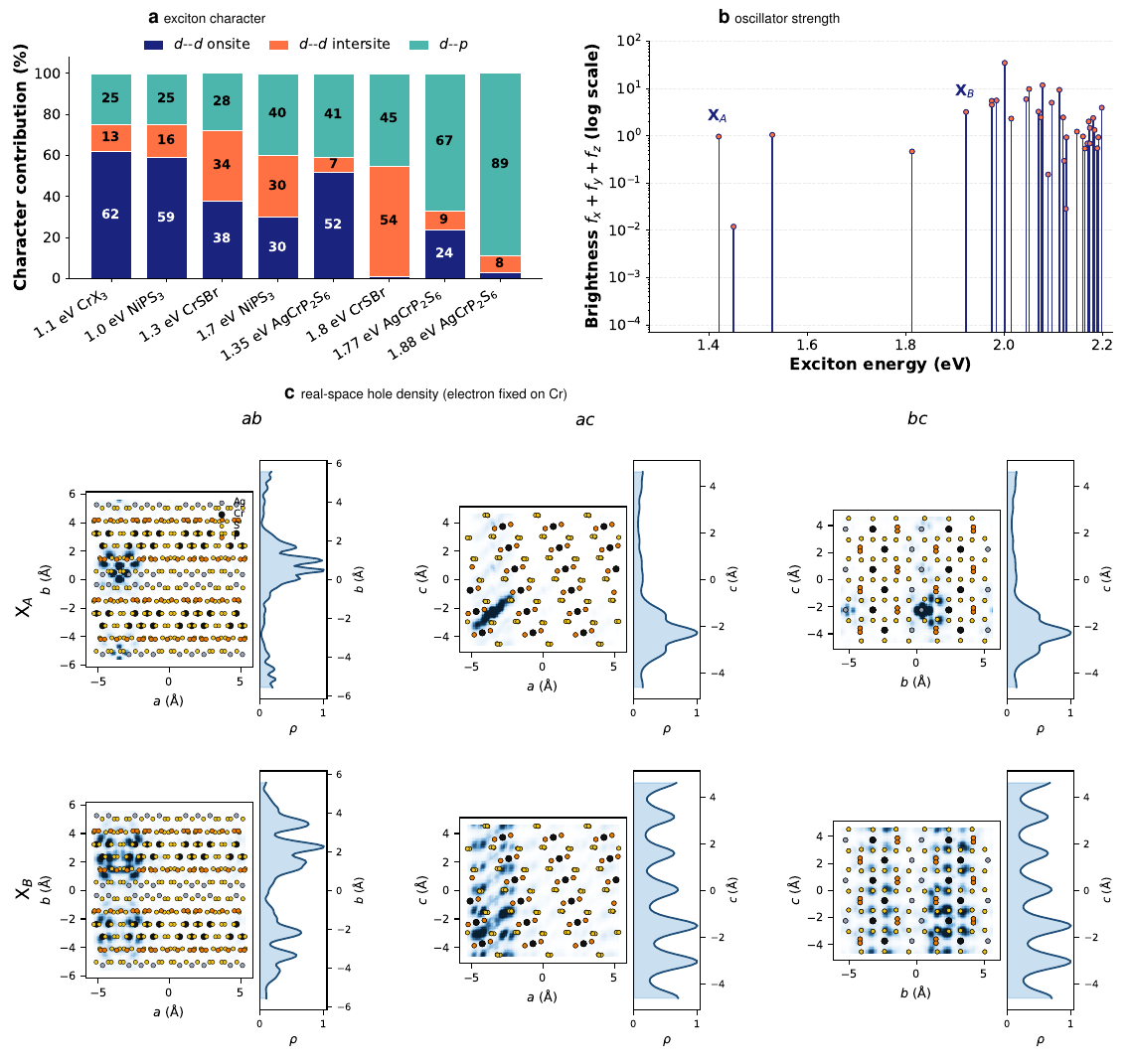}
\caption{Exciton character, oscillator strength, and photoluminescence. (a) Character composition for representative low-energy excitations in CrX$_3$, NiPS$_3$, CrSBr, and AgCrP$_2$S$_6$, sorted by total $d$--$d$ weight. (b) Total dipole brightness $f_x+f_y+f_z$ on a logarithmic scale for every BSE eigenstate between 1.3 and 2.2 eV; X$_A$ marks the lowest bright state at 1.42 eV and X$_B$ the representative ligand-assisted state near 1.88 eV. (c) Real-space hole density for X$_A$ and X$_B$ with the electron fixed on a Cr site. Rows: X$_A$ and X$_B$. Columns: $ab$, $ac$, and $bc$ projections, each with the corresponding one-dimensional density $\rho$ along the vertical axis of the map. X$_A$ remains tightly localized about a single Cr-centered ligand cage, whereas X$_B$ is more extended within the layer while still lacking coherent interlayer delocalization.}
%Red, yellow, dark blue, and light blue atoms in (c) represent P, S, Cr, and Ag, respectively.}
\label{fig:exciton-character}
\end{figure*}

The excitonic consequences of that competition are summarized in Fig.~\ref{fig:exciton-character}. In the language commonly used for local ligand-field~\cite{Sugano1954,SuganoBook} and charge-transfer excitations~\cite{Zhang1988,Lee2006,RevModPhys.83.705,ghosh2023magnetic}, the lowest excitons in AgCrP$_2$S$_6$ retain a large Frenkel character. The lowest computed bright excitonic state at 1.42 eV (experimentally it comes out slightly lower in energy $\sim$1.35 eV) has a strong binding energy $\sim$770 meV and retains 52\% onsite $d$--$d$ character and only 7\% intersite $d$--$d$ weight, even though 41\% of its amplitude already involves ligand-assisted $d$--$p$ mixing [Figs.~\ref{fig:exciton-character}(a)]. Higher excitons progressively pick up more ligand weight, but the intersite $d$--$d$ contribution remains modest compared with the more delocalized response seen in other ferromagnetic Cr-based analogues.

\begin{figure*}[t]
\centering
\includegraphics[width=0.98\linewidth]{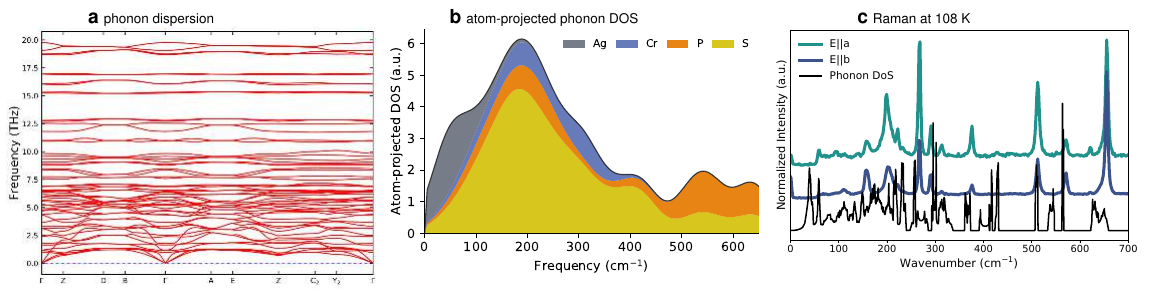}
\caption{Lattice dynamics of AgCrP$_2$S$_6$. (a) Phonon dispersion of the optimized structure, showing no unstable modes along the selected high-symmetry path. (b) Atom-projected phonon density of states. (c) Polarization-resolved Raman spectra at 108\,K compared with the calculated phonon density of states (298\,K data in the Supplementary Information).}
\label{fig:phonon-raman}
\end{figure*}

In CrX$_3$, low-energy excitons are local but still benefit from ferromagnetic spin alignment, which lets the electron and hole hop through neighboring Cr sites of the same spin~\cite{acharya2022real}. In CrSBr, low symmetry (orthorhombic) coexists with a strong and highly anisotropic excitonic response, showing that reduced symmetry alone does not force optical weakness~\cite{bianchi2023paramagnetic,watson2024giant,ruta2023hyperbolic,shao2024exciton}. In AgCrP$_2$S$_6$ two separate effects keep the lowest exciton local. First, the Cr network is antiferromagnetic. Hopping of an electron or hole onto a neighbor then meets the opposite spin and costs exchange energy, as in the classic AFM carrier problem~\cite{BrinkmanRice1970,SchmittRink1988}. The same loss of kinetic coherence appears when a ferromagnetic chain is disordered~\cite{ruta2023hyperbolic,watson2024giant}. Second, Ag$^+$ is a filled $d^{10}$ shell. A Cr$\to$Ag intersite $d$--$d$ path has nowhere to put the electron in the Ag $d$ manifold, so that channel is blocked. Ligand hybridization still narrows the gap, yet the lowest BSE exciton keeps only a few percent intersite $d$--$d$ weight. The result is not a purely atomic excitation: ligand hybridization remains substantial, as seen from the real-space densities in Fig. \ref{fig:exciton-character}. Instead, the material realizes a mixed but still strongly localized excitonic regime in which $d$--$p$ admixture narrows the gap without restoring large intersite $d$--$d$ transport. This balance explains why AgCrP$_2$S$_6$ remains more localized than CrSBr in the exciton decomposition, despite its covalent sulfur-phosphorus framework.

The polarization-resolved solution of the Bethe-Salpeter equations (BSE) further constrains the assignment of the lowest exciton clusters. The first theoretical state (X$_A$) in that manifold appears at 1.42 eV and has oscillator strengths $(f_x,f_y,f_z)=(0.03,<10^{-5},0.9)$ in Cartesian coordinates, placing nearly all of its optical weight along $z$ [Fig.~\ref{fig:exciton-character}(b)]. The nearby 1.45 eV state is much weaker and more mixed, whereas the 1.53 eV state is predominantly $x$ polarized. The representative ligand-assisted state X$_B$ near 1.88 eV shows oscillator strengths $(f_x,f_y,f_z)=(0.075,1.0\times10^{-4},1.26)$, i.e.~$(5.6\%,<10^{-2}\%,94.4\%)$ in fractional weight, so it is again strongly $z$ polarized and shares the same $B_u$ symmetry as X$_A$. 

In the conventional $C_{2h}$ point symmetry group for monoclinic AgCrP$_2$S$_6$, in which the twofold rotation axis lies along the crystallographic $b$ direction, the $x$ and $z$ Cartesian dipoles (within the $ac$ plane) transform as $B_u$ while the $y$ dipole (along $b$) transforms as $A_u$; the calculated lowest bright state at 1.42 eV is therefore best viewed as a predominantly $B_u$, optically allowed Frenkel exciton, and the higher-lying X$_B$ at 1.88 eV is a more ligand-extended state of the same irreducible representation. Excitons transforming as $A_u$ are dark for the dominant $z$ polarization of X$_A$ in the ideal crystal. $B_g$ phonons can mix $B_u$ and $A_u$ character and thereby borrow intensity. Fixing the electron at a Cr site reveals a marked contrast in the real-space hole distributions of $X_A$ and $X_B$ [Fig.~\ref{fig:exciton-character}(c)]. While the lowest-lying bright state $X_A$ exhibits a tightly localized, Frenkel-like hole density confined to a single Cr-centered ligand cage, $X_B$ displays substantial spatial extension within the monolayer $ab$-plane while maintaining a strong out-of-plane confinement along $c$. 
%The present results are consistent with a weakly bright, predominantly non-spin-flip local exciton rather than a spin-flip excitation whose intensity is borrowed only indirectly~\cite{acharyaTheoryColorsStrongly2023}. 

Relative to the intense low-energy excitons discussed in CrSBr and related anisotropic magnets, the AgCrP$_2$S$_6$ onset is therefore better described as a weak but symmetry-selective $q=0$ excitation. The excitonic weights we access from our computed eigenfunctions within the BSE solutions within $\mathrm{QS}G\hat{W}$ framework also do not reveal a clean sequence of bright-dark partner states built from alternating symmetric and antisymmetric band combinations; instead, the low-energy spectral weight stays concentrated in a small set of diagonal channels, while the oscillator strength changes mainly through polarization anisotropy and overall suppression. For strongly localized $d$--$d$/$d$--$p$ excitons, that anisotropy means that the lattice primarily couples by modulating the local ligand field and covalency and by borrowing optical weight through symmetry lowering, rather than through the nonlocal bond-modulation channel that becomes effective when intersite $d$--$d$ hopping is large~\cite{RevModPhys.83.705,Devereaux2007,Hancock2010,acharyaTheoryColorsStrongly2023}.

\subsection*{Exciton-Phonon Coupling in \texorpdfstring{AgCrP$_2$S$_6$}{AgCrP2S6}}

\subsubsection*{Symmetry-Based Analysis of Coupling Channels}

\begin{table*}[t]
\caption{Symmetry-based exciton-phonon coupling channels for the low-energy excitons of AgCrP$_2$S$_6$. The oscillator weights are taken from the BSE output in Cartesian coordinates. The phonon candidates are restricted to zone-center modes relevant to the Raman comparison in Fig.~\ref{fig:phonon-raman}. The odd-parity $A_u$/$B_u$ phonons near 328.0 and 366.8 cm$^{-1}$ are first-order Raman silent in the ideal $C_{2h}$ crystal and are therefore not included as direct vibronic partners; they can appear only as disorder-activated channels through Ag-site or other inversion-breaking fields.}
\label{tab:exciton-phonon-symmetry}
\small
\begin{ruledtabular}
\begin{tabular}{@{}llll@{}}
\parbox[t]{0.17\textwidth}{\raggedright \textbf{Exciton}} &
\parbox[t]{0.22\textwidth}{\raggedright \textbf{Dominant optical channel}} &
\parbox[t]{0.25\textwidth}{\raggedright \textbf{Symmetry consequence}} &
\parbox[t]{0.24\textwidth}{\raggedright \textbf{Leading phonon partners}} \\
\parbox[t]{0.17\textwidth}{\raggedright 1.42 eV (theory, X$_{A}$)} &
\parbox[t]{0.22\textwidth}{\raggedright $z$ polarized, $(3\%,<0.1\%,97\%)$, predominantly $B_u$} &
\parbox[t]{0.25\textwidth}{\raggedright $A_g$ modes give diagonal energy renormalization; $B_g$ modes mix bright $B_u$ and dark $A_u$ channels} &
\parbox[t]{0.24\textwidth}{\raggedright $A_g$: 279.4, 302.9 cm$^{-1}$; $B_g$: 292.6, 310.8, 364.7 cm$^{-1}$} \\
\parbox[t]{0.17\textwidth}{\raggedright 1.45 eV (theory)} &
\parbox[t]{0.22\textwidth}{\raggedright weak mixed polarization} &
\parbox[t]{0.25\textwidth}{\raggedright sensitive to symmetry lowering and oscillator-strength borrowing} &
\parbox[t]{0.24\textwidth}{\raggedright secondary partner in $B_g$-assisted mixing} \\
\parbox[t]{0.17\textwidth}{\raggedright 1.53 eV (theory)} &
\parbox[t]{0.22\textwidth}{\raggedright $x$ polarized, $(97\%,0.2\%,3\%)$, also $B_u$-like} &
\parbox[t]{0.25\textwidth}{\raggedright shares the same bright manifold as the $z$-polarized state but with orthogonal in-plane anisotropy} &
\parbox[t]{0.24\textwidth}{\raggedright can participate in higher-energy $A_g/B_g$ vibronic redistribution} \\
\parbox[t]{0.17\textwidth}{\raggedright 1.88 eV (theory, X$_B$)} &
\parbox[t]{0.22\textwidth}{\raggedright $z$ polarized, $(5.6\%,<10^{-2}\%,94.4\%)$, predominantly $B_u$ with substantial ligand-$p$ admixture} &
\parbox[t]{0.25\textwidth}{\raggedright same selection rules as the 1.42 eV state ($A_g$ diagonal, $B_g$ off-diagonal to $A_u$), but with coupling weight shifted toward ligand-bond modes that modulate the Cr-S/P-S covalency} &
\parbox[t]{0.24\textwidth}{\raggedright same $A_g$/$B_g$ classes as X$_A$; ligand-dominated $A_g$ stretching modes provide the diagonal channel, $B_g$ modes mediate bright-dark mixing within the higher-lying manifold} \\
\end{tabular}
\end{ruledtabular}
\end{table*}

Given the localized $d$--$d$/$d$--$p$ character of excitons in AgCrP$_2$S$_6$, one would expect relatively strong coupling to the lattice. To probe the role of exciton-phonon coupling in this system, we first computed and measured the phonon spectrum. Figure~\ref{fig:phonon-raman} compares the calculated vibrational spectrum [Fig.~\ref{fig:phonon-raman}(a)] and atom-projected phonon DOS [Fig.~\ref{fig:phonon-raman}(b)] with the Raman measurements taken at 108\,K [Fig.~\ref{fig:phonon-raman}(c)] (298\,K spectra in the Supplementary Information). The calculated phonon dispersion has no unstable branches, verifying the dynamical stability of the optimized structure, and the main Raman features track the calculated vibrational density of states in both in-plane polarizations. The atom projection shows that the mid-frequency window relevant to the PL sidebands (vide infra) carries substantial S (and Cr) weight, while the highest-frequency stretches are dominated by P/S motion, consistent with assigning X$_A$ partners to more local Cr--S crystal-field modes and X$_B$ partners to ligand-bond modes.

The 20-atom unit cell ($Z=4$ of Ag$_{0.5}$Cr$_{0.5}$PS$_3$) contributes 60 zone-center modes that decompose as $\Gamma_\mathrm{opt} = 14A_g + 16B_g + 13A_u + 14B_u$ (acoustic: $A_u + 2B_u$), providing 30 Raman-active and 27 IR-active optical modes; all symmetry classes are represented in the measured and calculated spectra. A few experimental Raman peaks appear slightly higher in energy than the nearest calculated features, while others align closely; the overall match is sufficient to identify the active symmetry windows and to rule out a large structural reorganization between the crystallographic model and the vibrational response. Absolute frequencies still differ by a few tens of cm$^{-1}$. The phonons are computed for a DFT-relaxed, fully ordered Ag/Cr stripe geometry (a $T=0$ ordered DFT structure initialized from the 100\,K $P2/c$ refinement), while the Raman spectra are measured on crystals that retain weak low-temperature Ag disorder, which can shift absolute peak positions. Semilocal DFT also misses part of the magnetic-correlation physics that stretches metal--ligand bonds, so the relaxed DFT bond lengths and force constants remain slightly mistuned relative to experiment and move the phonon energies on this scale. Auti \textit{et al.}~\cite{Auti_Antipolar} likewise used finite-displacement Phonopy DFT-D3 phonons on an AFM-relaxed AgCrP$_2$S$_6$ cell; their measured Raman peaks near $269$, $312$, and $376$\,cm$^{-1}$ agree with ours to within $\sim$2\,cm$^{-1}$ [Table~\ref{tab:raman-pl-modes}], and the residual theory--experiment offsets are of the same few-tens-of-cm$^{-1}$ size. We use the calculations to assign symmetries and mode character, and the measured peaks for absolute positions. This experimental-theoretical consistency establishes that the low-energy optical response occurs in a dynamically stable monoclinic lattice with a local Ag disorder, rather than near a soft-mode instability.
%AgCrP$_2$S$_6$ therefore differs from cases where lattice fluctuations are the dominant localization channel~\cite{tise2}.

Next, we computed the symmetry-based exciton-phonon coupling channels for the low-energy excitons of AgCrP$_2$S$_6$. The results are tabulated in Table~\ref{tab:exciton-phonon-symmetry}. The symmetry classification follows directly from the $C_{2h}$ selection rule: the exciton--phonon coupling element $\langle\psi_i|\hat{H}_\mathrm{ep}|\psi_j\rangle$ is nonzero only if $\Gamma_i \otimes \Gamma_\mathrm{ph} \otimes \Gamma_j \supset A_g$. For the diagonal channel $\langle B_u|\hat{H}_\mathrm{ep}|B_u\rangle$, the direct product $B_u \otimes B_u = A_g$ requires $\Gamma_\mathrm{ph} = A_g$. For the off-diagonal channel $\langle B_u|\hat{H}_\mathrm{ep}|A_u\rangle$, one finds $B_u \otimes A_u = B_g$, which requires $\Gamma_\mathrm{ph} = B_g$. Totally symmetric $A_g$ phonons therefore provide the diagonal energy renormalization and Franck-Condon broadening channel for a $B_u$ exciton~\cite{Devereaux2007,Hancock2010}, while $B_g$ phonons are the leading Raman-active modes that can linearly mix a bright $B_u$ state with nearby $A_u$ dark states and thereby redistribute oscillator strength or rotate the polarization anisotropy without requiring a structural phase transition~\cite{RevModPhys.83.705}. The $C_{2h}$ rules fix which phonon symmetries can couple. Which modes carry the Franck--Condon weight is read from the local $d$--$d$/$d$--$p$ character of X$_A$, the mid-frequency Raman window with Cr--S weight, and the PL sidebands below.

Allowing for the modest frequency offsets between calculated and measured Raman peaks visible in Fig.~\ref{fig:phonon-raman}, this symmetry filter identifies the $B_g$ modes at 292.6, 310.8, and 364.7 cm$^{-1}$ and the $A_g$ modes at 279.4 and 302.9 cm$^{-1}$ as the most natural vibronic partners of the lowest energy (1.42 eV) $B_u$ exciton (X$_A$). The odd-parity $A_u/B_u$ modes are less natural Raman assignments in an ideal crystal, but they remain plausible only to the extent that local Ag disorder or other inversion-breaking fields in a non-ideal crystal could activate them. This hierarchy is consistent with the exciton decomposition itself: the 52\% onsite $d$--$d$ and 41\% ligand-assisted $d$--$p$ weights favor local Cr-S crystal-field and covalency fluctuations, whereas the very small 7\% intersite $d$--$d$ component disfavors a predominantly intersite coupling mechanism.

The higher-lying ligand-assisted exciton X$_B$ near 1.88 eV belongs to the same $B_u$ representation as X$_A$ and is therefore subject to identical selection rules: $A_g$ phonons drive its diagonal Franck-Condon renormalization and $B_g$ phonons mediate any linear mixing with nearby $A_u$ dark states. What changes between X$_A$ and X$_B$ is not the symmetry channel but the spectral weight of the modes that contribute to each channel. Because X$_B$ shows substantially larger ligand-$p$ admixture and a clearly more extended real-space density within the thiophosphate layer [Fig.~\ref{fig:exciton-character}(c)], its coupling is expected to be carried preferentially by $A_g$ and $B_g$ modes that modulate the Cr-S and P-S bond lengths rather than by the more local Cr-S crystal-field modes that dominate X$_A$. In the calculated zone-center spectrum, these are the totally symmetric P-S and S-S stretching modes in the upper Raman window, which couple naturally to a ligand-extended $B_u$ exciton. In a polarization-resolved Raman or resonance-Raman experiment, the two excitons should therefore behave as $B_u$ partners of the same selection rules, but with X$_A$ resonantly enhancing the lower-frequency Cr-S crystal-field $A_g$/$B_g$ modes and X$_B$ resonantly enhancing the higher-frequency ligand-bond $A_g$/$B_g$ modes.

\begin{figure*}[t]
\centering
\includegraphics[width=0.98\linewidth]{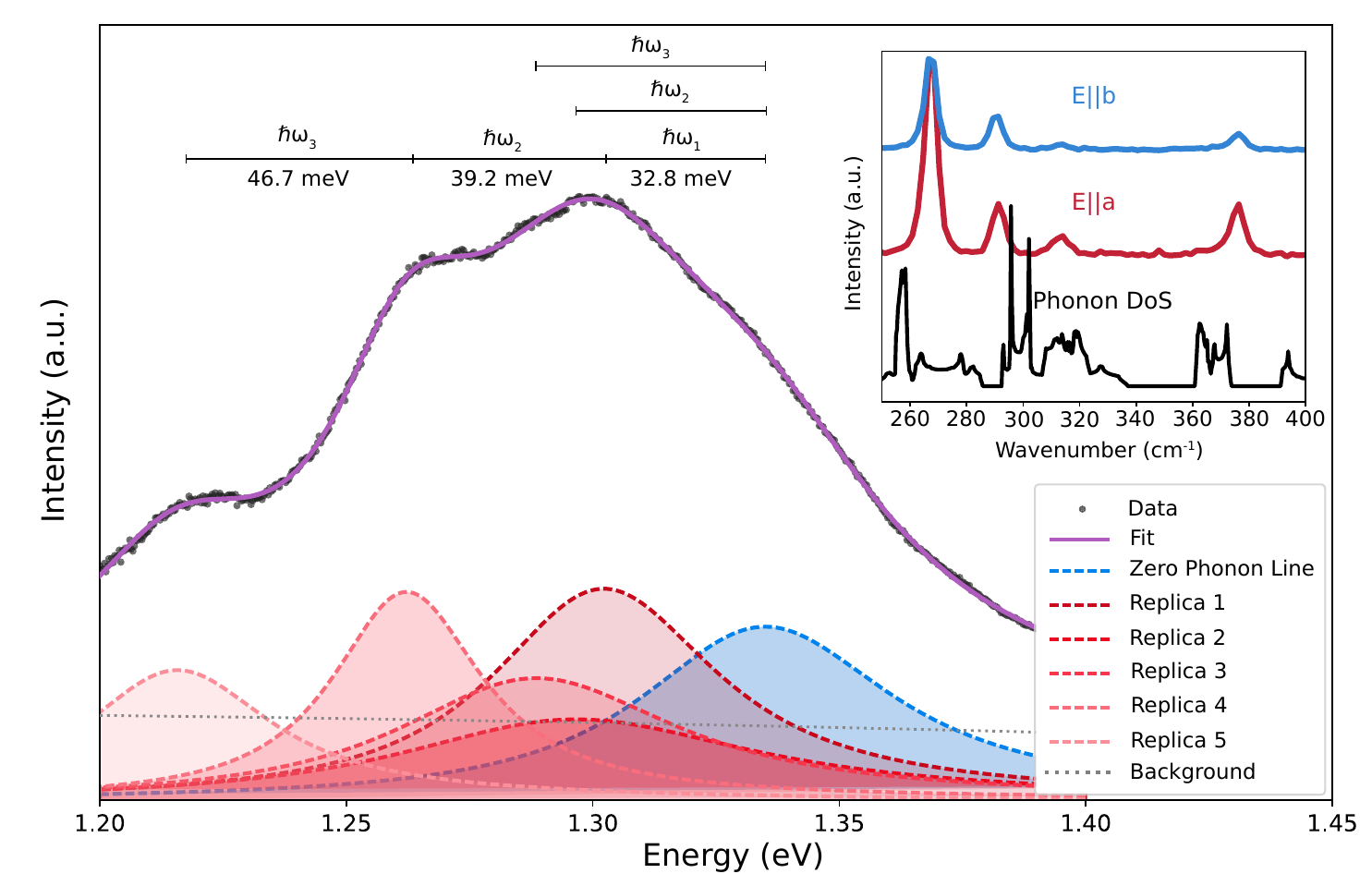}
\caption{Photoluminescence at 5\,K with a multi-Voigt fit; replicas 1--5 are phonon sidebands of the X$_A$ recombination. }
\label{fig:pl}
\end{figure*}

\subsubsection*{Experimental Evidence for Exciton-Phonon Coupling}

We find compelling experimental evidence for the calculated exciton-phonon coupling in the photoluminescence spectrum of AgCrP$_2$S$_6$ [Fig.~\ref{fig:pl}], taken on a AgCrP$_2$S$_6$ single crystal at 5\,K.

The PL spectrum consists of a relatively broad envelope with multiple sub-peaks, resembling a typical Franck-Condon progression of vibronic emission features. The high energy broad shoulder occurs at an energy of 1.33 eV, consistent with previous studies of photoluminescence in AgCrP$_2$S$_6$ \cite{park_anisotropic_2024, volochanskyi_indirect_2026}. Since this is the highest-energy feature identified in the spectrum, the shoulder at 1.33 eV is assigned as the zero-phonon line (ZPL), corresponding to radiative recombination of the  lowest-energy excitonic state, or the $X_A$ exciton, in AgCrP$_2$S$_6$.  Within a Franck-Condon framework, the five satellite peaks on the low-energy side of the ZPL represent phonon sidebands. 

To determine the primary phonon modes participating in the recombination, and the effective exciton-phonon coupling strength exemplified in this transition, we employed a three-mode Huang-Rhys model. In the Franck-Condon picture of exciton-phonon coupling, the spectral function follows a Poisson distribution \cite{Spano_MultiMode}, given by

\begin{equation}
A(E) = \mathrm{e}^{-S} \sum_{n=0}^\infty \frac{S^n}{n!}L(E -E_0 + n\hbar\omega_{ph}),
\end{equation}

where $S$ is the Huang-Rhys factor which corresponds to the strength of exciton-phonon coupling, $n$ refers to the number of emitted phonons with frequency $\omega_{ph}$, $L$ is a line shape function of emission, and $E_0$ is the energy of the zero phonon line. In considering a system where multiple phonon modes couple to the excitonic state, one can fit the photoluminescence spectrum with multiple Huang-Rhys factors \cite{Spano_MultiMode, Spano_multimode_PL}, where the fit includes the product of multiple Franck-Condon factors. In the case of a three-mode fit, we apply the model

\begin{equation}
I_{HR}(E) = I_0 \mathrm{e}^{-(S_1 + S_2 + S_3)} \sum_{n_1=0}^\infty\sum_{n_2=0}^\infty\sum_{n_3=0}^\infty\frac{S_1^{n_1}}{n_1!}\frac{S_2^{n_2}}{n_2!}\frac{S_3^{n_3}}{n_3!}V(E; E_0 - n_1\hbar\omega_1 - n_2\hbar\omega_2 -n_3\hbar\omega_3, \sigma_n, \gamma_n),
\label{three_mode_HR}
\end{equation}

where $I_{HR}$ refers to the intensity of the phonon replicas; $I_0$ is an intensity normalization factor such that $I_0 \mathrm{e}^{-(S_1 + S_2 + S_3)}$ is the intensity of the zero-phonon line; $S_1$ $S_2$, and $S_3$ refer to the Huang-Rhys factors for phonon modes $\omega_1$, $\omega_2$, and $\omega_3$; $n_1$,  $n_2$, and $n_3$ refer to the integer number of emitted phonons with frequencies $\omega_1$, $\omega_2$, and $\omega_3$, respectively; $V(E; E_c, \sigma, \gamma)$ is an area-normalized Voigt profile centered at an energy of $E_c$ with broadening described by $\sigma$ and $\gamma$; and $E_0 = 1.33$ eV is the energy of the zero phonon line.

The sum in Eq.~\eqref{three_mode_HR} is truncated to the replicas resolved in the PL and expected from the leading Raman partners of $X_A$. Single-, two-, and three-mode progressions are compared in the Supplementary Information; the three-mode model in Fig.~\ref{fig:pl} is the one retained in the main text, with phonon energies that match the measured Raman peaks.

The values (n$_{1}$, n$_{2}$, n$_{3}$) as shown in equation \ref{three_mode_HR} were fitted to the combinations (1, 0, 0), (0, 1, 0), (0, 0, 1), (1, 1, 0), and (1, 1, 1) for replicas 1, 2, 3, 4, and 5, respectively. Consequently, the truncated Huang--Rhys progression used here is
\begin{equation}
\begin{aligned}
I_{\mathrm{HR}}(E)
&=
I_0 e^{-(S_1+S_2 + S_3)}
\Big[
V(E;E_0,\sigma_0,\gamma_0) \\
&\quad+
S_1V(E;E_0-\hbar\omega_1,\sigma_1,\gamma_1) \\
&\quad+
S_2V(E;E_0-\hbar\omega_2,\sigma_2,\gamma_2) \\
&\quad+
S_3V(E;E_0-\hbar\omega_3,\sigma_3,\gamma_3) \\
&\quad+
S_1S_2V(E;E_0-\hbar\omega_1-\hbar\omega_2,\sigma_4,\gamma_4) \\
&\quad+
S_1S_2S_3V(E;E_0-\hbar\omega_1-\hbar\omega_2 - \hbar\omega_3,\sigma_5,\gamma_5)\Big].
\end{aligned}
\end{equation}

Here, $V(E;E_c,\sigma_j,\gamma_j)$ is an area-normalized Voigt profile centered at energy $E_c$, where $\sigma_j$ and $\gamma_j$ describe the Gaussian and Lorentzian broadening, respectively, of spectral component $j$. The index $j=0$ denotes the zero phonon line, while $j=1,\ldots,5$ identifies the phonon replicas.

Following this fitting procedure, we determine the three resolved phonon modes that contribute to the vibronic progression to be $\hbar\omega_1$ = 32.79 meV ($\nu_1$ = 264.5 cm\textsuperscript{-1}), $\hbar\omega_2$ = 39.24 meV ($\nu_2$ = 316.5 cm\textsuperscript{-1}), and $\hbar\omega_3$ = 46.74 meV ($\nu_3$ = 376.9 cm\textsuperscript{-1}), with effective Huang-Rhys factors of $S_1$ = 1.210, $S_2$ = 0.695, and $S_3$ = 0.859, respectively. This fit thus yields an effective Huang-Rhys factor of $S_{\textrm{total}} = \sum_1^3S_i = 2.764$, with a 95\% confidence interval of $[2.616, 3.136]$.

\begin{table*}[t]
\caption{Raman modes associated with the $X_A$ exciton in AgCrP$_2$S$_6$. Raman modes in the first column are obtained from a measurement at 108 K in this study, while cited experimental modes reference data obtained at 6 K and 5 K. Quoted uncertainties for this study are determined as half the difference of $E||a$ and $E||b$ orientations.}
\label{tab:raman-pl-modes}
\small
\begin{ruledtabular}
\begin{tabular}{@{}lllll@{}}
\parbox[t]{0.1\textwidth}{\raggedright \textbf{Mode}} &
\parbox[t]{0.2\textwidth}{\raggedright \textbf{Raman Measurement, This Study (cm$^{-1}$)}} &
\parbox[t]{0.21\textwidth}{\raggedright \textbf{Raman Measurements, Previous Studies (cm$^{-1}$)}} &
\parbox[t]{0.15\textwidth}{\raggedright \textbf{Calculated Mode (cm$^{-1}$)}} & 
\parbox[t]{0.25\textwidth}{\raggedright \textbf{Photoluminescence-Derived Phonon Mode (cm$^{-1}$)}} \\ \addlinespace \hline \addlinespace
\parbox[t]{0.1\textwidth}{\raggedright $\nu_1$} &
\parbox[t]{0.2\textwidth}{\raggedright 267.5 $\pm$ 0.1} &
\parbox[t]{0.21\textwidth}{\raggedright 269.3 $\pm$ 0.0 \cite{Auti_Antipolar}; 266.3 \cite{Xu_Mirror}} &
\parbox[t]{0.15\textwidth}{\raggedright 279.4} &
\parbox[t]{0.25\textwidth}{\raggedright 264.5} \\
\parbox[t]{0.1\textwidth}{\raggedright $\nu_2$} &
\parbox[t]{0.2\textwidth}{\raggedright 313.5} &
\parbox[t]{0.21\textwidth}{\raggedright 312.1 $\pm$ 0.3 \cite{Auti_Antipolar}} &
\parbox[t]{0.15\textwidth}{\raggedright 310.8} &
\parbox[t]{0.25\textwidth}{\raggedright 316.5} \\
\parbox[t]{0.1\textwidth}{\raggedright $\nu_3$} &
\parbox[t]{0.2\textwidth}{\raggedright 376.3 $\pm$ 0.1} &
\parbox[t]{0.21\textwidth}{\raggedright 376.1 $\pm$ 0.3 \cite{Auti_Antipolar}} &
\parbox[t]{0.15\textwidth}{\raggedright 364.7; 366.8} &
\parbox[t]{0.25\textwidth}{\raggedright 376.9} \\
\end{tabular}
\end{ruledtabular}
\end{table*}

%\Jessica comments:
%\textit{Things I want to say in the discussion here:
%\begin{itemize}
%\item In general, the PL spectrum shows evidence of the XA exciton coupling to vibrational modes in the range of 265-375 cm-1 in its recombination, as the theory suggests \\
%\item Discussion of the modes in particular: 266 lines up very well with the dominant mode seen in the Raman spectra, 301 is in between observed modes and aligns with theory mode, 366 lines up well with calculated and is similar to theory \\
%\item The HR factor indicates moderate to strong coupling. Compare to measurements in other 2D magnets and maybe TMDCs. 
%\item The moderate to strong coupling supports the localized character of %the excitons
%\end{itemize}
%}
Taken together, the PL sidebands place the $X_A$ recombination in the mid-frequency vibrational window of roughly $265$--$377$\,cm$^{-1}$, precisely the range identified by the symmetry analysis as the natural $A_g$/$B_g$ partner space of the lowest $B_u$ exciton [Table~\ref{tab:exciton-phonon-symmetry} and Fig.~\ref{fig:phonon-raman}]. Within that window, the three resolved Franck--Condon modes align closely with independent Raman and calculated frequencies [Table~\ref{tab:raman-pl-modes}]: the $264.5$\,cm$^{-1}$ replica coincides with the dominant Raman peak measured here at $267.5$\,cm$^{-1}$ (and with literature values near $266$--$269$\,cm$^{-1}$); the $316.5$\,cm$^{-1}$ replica coincides with the Raman peak measured at $313.5$\,cm$^{-1}$ and matches the calculated $A_g$ mode at $310.8$\,cm$^{-1}$; and the $376.9$\,cm$^{-1}$ replica aligns with the calculated $364.7$/$366.8$\,cm$^{-1}$ pair while agreeing well with the experimental Raman window near $376$\,cm$^{-1}$. 

%The corresponding Huang--Rhys parameter ($S = g^2$, where $g$ is the dimensionless exciton--phonon coupling constant) indicates moderate-to-strong coupling. This is substantially larger than the weak coupling ($S \ll 1$) typical of delocalized free excitons in conventional transition-metal dichalcogenides, comparable to the strong polaronic/vibronic regime reported for $\mathrm{CrI}_3$ ($g \sim 1.5$, $S \sim 2.3$)~\cite{Jin2020CrI3Polaron}, and still well below the extreme $d\text{--}d$/phonon coupling inferred for $\mathrm{NiPS}_3$ ($g \sim 10$)~\cite{Ergecen2021NiPS3}. Such a Huang--Rhys strength is the expected spectroscopic counterpart of a spatially localized Frenkel-like excitation: a more coherent, intersite-delocalized exciton would couple more weakly to local $\mathrm{Cr\text{--}S}$ crystal-field modes and would not support a multi-mode Franck--Condon progression of this magnitude. The PL analysis therefore closes the loop between the BSE character/real-space decomposition of $X_A$ [Fig.~\ref{fig:exciton-character}] and the lattice dynamics of Fig.~\ref{fig:phonon-raman}.

The extracted Huang--Rhys parameter indicates moderate-to-strong exciton--phonon coupling in AgCrP$_2$S$_6$. This value stands in sharp contrast to the weak coupling regime ($S \ll 0.1$) characteristic of delocalized Wannier--Mott excitons in conventional transition-metal dichalcogenides (TMDs). Instead, it falls squarely into the strong vibronic regime observed in localized 2D systems, such as monolayer $\mathrm{CrI}_3$ ($S \approx 2.3$)~\cite{Jin2020CrI3Polaron} and 2D halide perovskites ($S \sim 1\text{--}3$). A Huang--Rhys parameter of this magnitude is the direct spectroscopic signature of a spatially compact, Frenkel-like excitation: upon photoexcitation, the localized exciton induces a strong local lattice distortion, driving a pronounced multi-phonon Franck--Condon progression. This photoluminescence analysis directly connects the real-space localization of $X_A$ derived from the BSE calculations [Fig.~\ref{fig:exciton-character}] to the strong lattice dynamics observed in Raman spectroscopy [Fig.~\ref{fig:phonon-raman}].

\section*{Discussion}

AgCrP$_2$S$_6$ combines a covalent low-symmetry thiophosphate framework with an antiferromagnetic Cr sublattice, where the ordered Ag/Cr stripe motif and filled-shell Ag$^+$ sites restrict intersite $d$--$d$ paths relative to other magnetic Cr-based compounds. Diffraction shows that the monoclinic structure persists from 100 to 300\,K, with additional low-temperature Ag disorder, while the many-body calculations show that the reduced quasiparticle gap and higher covalency do not produce strongly delocalized low-energy excitons. The lowest bright excitation retains a strong onsite $d$--$d$ weight, with a nearly comparable ligand mixed $d$--$p$ character that increases at higher-energy excitonic transitions, as established by the exciton decomposition, polarization-resolved oscillator strengths, and real-space hole densities of X$_A$ and X$_B$.

Symmetry analysis identifies the calculated onset near 1.4\,eV as a strongly anisotropic, predominantly $B_u$ Frenkel exciton, whose lattice coupling is governed by $A_g$ diagonal renormalization and $B_g$-mediated vibronic mixing. Photoluminescence at 5\,K confirms that picture experimentally: the X$_A$ zero-phonon line supports a Franck--Condon progression built from mid-frequency modes near $265$, $316$, and $377$\,cm$^{-1}$, matching the Raman/calculated phonon window, with an effective Huang--Rhys factor $S_{\mathrm{eff}}=2.764$ that signals moderate-to-strong exciton--phonon coupling and reinforces the localized character of the recombination. AgCrP$_2$S$_6$ therefore realizes a regime in which covalency, low symmetry, and antiferromagnetic exchange are all important, yet the low-energy excitation remains local because AFM order restricts electron and hole hopping on the Cr network, and Ag$^+$ is a filled $d^{10}$ shell that blocks Cr$\to$Ag $d$--$d$ paths. The same hierarchy of local exciton character, selective $A_g$/$B_g$ vibronic partners, and a measurable Franck--Condon progression provides concrete targets for polarization-resolved Raman spectroscopy, resonance Raman spectroscopy, and absorption anisotropy on oriented crystals. More broadly, contrasting AgCrP$_2$S$_6$ with Cr-, Ni-, and Mn-based thiophosphates and chromium trihalides of differing local symmetry and crystal field establishes cation substitution and ligand-ion engineering as a transferable design principle for tuning the degree of exciton localization across this family of layered magnets, with direct implications for exciton-boson (phonon and magnon) coupling and for spintronic functionality.

\section*{Methods}
\subsubsection{Theoretical Methods}

The electronic structure and the corresponding excitonic properties of bulk AgCrP$_2$S$_6$ are computed within the self-consistent \textit{ab initio} many-body perturbative framework QS$G\hat{W}$~\cite{cunningham2018,Cunningham2023}. QS$G\hat{W}$ is a self-consistent extension of quasiparticle self-consistent $GW$, QS$GW$, in the tradition developed by van Schilfgaarde, Kotani, and co-workers~\cite{qsgw,questaal_paper}. Within QS$G\hat{W}$ the electronic eigenfunctions are computed in the presence of screened Coulomb correlations corrected by an excitonic vertex and ladder diagrams in $W$~\cite{cunningham2018,Cunningham2023}. Self-consistency is imposed for both the self-energy $\Sigma$ and the charge density~\cite{Vidal10}. The latter is often neglected in standard $GW$, but it has been shown to qualitatively modify the electronic structure for several material classes~\cite{acharya2021importance,tise2}; it is particularly important when magnetic degrees of freedom are involved because spin polarization and $\Sigma$ are coupled.

For AgCrP$_2$S$_6$ we use the experimentally refined monoclinic bulk structure and the antiferromagnetic Cr arrangement discussed in the main text. The single-particle LDA calculations and the static quasiparticle calculations with QSGW and QS$G\hat{W}$ are performed with the Questaal package~\cite{questaal_paper,variational,Cunningham2023}. The static quasiparticlized self-energy $\Sigma^0(\mathbf{k})$ is generated on an $11\times7\times11$ mesh, while the relatively smoother dynamical self-energy is constructed on a $5\times3\times5$ mesh. Charge density and $\Sigma^0(\mathbf{k})$ are updated self-consistently until the root-mean-square change in the static self-energy falls below $2\times10^{-5}$ Ry.

Lattice-dynamical properties were computed using the displaced-supercell (frozen-phonon) approach with Phonopy~\cite{togo2023first} on a DFT-relaxed, fully ordered Ag/Cr stripe structure initialized from the experimental 100\,K $P2/c$ refinement, with van der Waals interlayer corrections included via the DFT-D3 scheme~\cite{grimme2010consistent}. Zone-center phonon frequencies and their irreducible representations under the $C_{2h}$ point group were obtained from that $T=0$ ordered DFT geometry.

The two-particle Hamiltonian used to compute both the vertex-corrected screening and the excitonic eigenvalues and eigenfunctions contains 60 valence bands and 30 conduction bands, which are sufficient to describe the Cr $3d$ states together with the relevant S/P ligand $p$ states in the low-energy optical window. This level of theory is needed because the central question is not only the size of the quasiparticle gap, but also whether the low-energy excitons remain local once covalency, antiferromagnetism, and low symmetry are treated on equal footing. Related self-consistent treatments of one- and two-particle spectra in ordered and disordered magnetic materials have been discussed in previous work~\cite{Cunningham2023,acharya2021electronic,acharya2022real,bianchi2023paramagnetic,watson2024giant}.

The polarization-resolved excitonic observables discussed in the main text are extracted from the same self-consistent QS$G\hat{W}$ workflow. In particular, the low-energy oscillator strengths identify the first strong $q=0$ onset as predominantly $z$ polarized, and the band/orbital decompositions show that the corresponding excitation remains dominated by diagonal, largely local transition channels. Those ingredients are what support the main conclusion of the paper: AgCrP$_2$S$_6$ realizes a low-symmetry covalent but still Frenkel-like limit, distinct from the more strongly coherent bright-exciton regimes established in Cr trihalides and CrSBr.

\subsubsection{Experimental Methods}
\label{sec:exp_methods}

\subsubsection*{Single-Crystal Synthesis}
Single crystals of $\mathrm{AgCrP_2S_6}$ were synthesized from elemental precursors in a 1:1:2:6 molar ratio, with a total charge of approximately 600~mg. The precursors were loaded into a quartz ampoule (14~mm inner diameter, 16~mm outer diameter) fitted with a 13~mm solid quartz plug, and evacuated to $\sim 80$~mTorr, and sealed under dynamic vacuum using a methane/oxygen torch, yielding a final reaction zone of $\sim 130$~mm. The ampoule was heated in a single-zone furnace to $400^{\circ}\mathrm{C}$ over 4~h, held for 4~h, ramped to $800^{\circ}\mathrm{C}$ over 4~h, held for 24~h, cooled to $750^{\circ}\mathrm{C}$ over 50~h, cooled to $700^{\circ}\mathrm{C}$ over 100~h, and maintained at $700^{\circ}\mathrm{C}$ for 100~h prior to furnace cooling to room temperature. The ampoules were opened, and the crystals were stored inside a nitrogen-filled glovebox to prevent ambient degradation.

\subsubsection*{Single-Crystal X-ray Diffraction}
Single-crystal XRD datasets were collected at 100~K and 300~K using a Bruker D8 Venture diffractometer equipped with a $\mathrm{Ga}$ MetalJet source ($\lambda = 1.341$~\AA). The crystals were mounted on a microloop using immersion oil. Data reduction and numerical absorption corrections were performed with \textsc{saint} and \textsc{sadabs}. The structures were solved using \textsc{shelxs}/\textsc{olex2} and refined via full-matrix least-squares on $|F|^2$ using \textsc{shelxl}~\cite{Krause2015,Dolomanov2009,Sheldrick2015}.

\subsubsection*{Raman Spectroscopy}
$\mathrm{AgCrP_2S_6}$ flakes were exfoliated onto $\mathrm{SiO_2/Si}$ substrates inside a nitrogen glovebox using Nitto tape (SWT-20) from single crystals epoxied to microscope slides. Substrates were mounted with thermal grease (Aremco Heat-Away$^{\mathrm{TM}}$ 641-EV) onto a Linkam THMS600 stage integrated into a Renishaw inVia Raman microscope. Excitation was provided by a 532~nm continuous-wave laser (Cobolt Samba) and dispersed using an 1800~lines/mm grating. The polarization state selection was controlled via integrated excitation optics, and the sample temperature was regulated via a liquid nitrogen flow. The peak centers were fitted using the standard Lorentzian profiles.

\subsubsection*{Cryogenic Photoluminescence Spectroscopy}
The PL spectra were acquired on a custom micro-photoluminescence setup. Bulk single crystals were secured to a 0.167~mm sapphire substrate (Valley Design) with Apiezon N-grease inside a glovebox, mounted onto a copper cold finger, and loaded into a closed-loop helium cryostation (Montana Instruments). Excitation was driven by a 488~nm continuous-wave laser (Melles Griot 35 LAS 450) focused through a $50\times$ objective ($\mathrm{NA} = 0.42$). Signal was collected through an 850~nm long-pass filter, fiber-coupled to a SpectraPro-2500i spectrograph (Acton Research), and detected using a liquid-nitrogen-cooled Si CCD (800~g/mm grating, 2~s exposure). The spectral intensity was calibrated using a NIST-traceable quartz tungsten halogen lamp (Teledyne Intellical QTHN0034) to generate a system responsivity curve.

\section*{Data availability}
CCDC 2583090-2583091 contains the supplementary crystallographic data produced experimentally and used in the theoretical calculations performed in this paper. The converged QS$G\hat{W}$ inputs and outputs needed to reproduce the electronic and excitonic results are available at Zenodo, DOI:\href{https://doi.org/10.5281/zenodo.23494277}{10.5281/zenodo.23494277}. 
%% AAB wants the CIF files to be put on CCDC
%During revisions, this text should be added in supporting information description section:
%CCDC 2583090-2583091contain the supplementary crystallographic data for this paper. These data can be obtained free of charge via www.ccdc.cam.ac.uk/data_request/cif, or by emailing data_request@ccdc.cam.ac.uk, or by contacting The Cambridge Crystallographic Data Centre, 12 Union Road, Cambridge CB2 1EZ, UK; fax: +44 1223 336033.

\section*{Code availability}
The electronic-structure and excitonic calculations were performed using the Questaal package~\cite{questaal_paper}, which is freely available at \url{https://www.questaal.org}.

\section*{Author contributions}
S.A.\ conceived and directed the study, performed the QS$G\hat{W}$ calculations, carried out the exciton decomposition and symmetry analysis, and led the manuscript preparation. D.P.\ implemented the key aspects of the QS$G\hat{W}$ and BSE methodology within the Questaal package and contributed to the calculations. M.v.S.\ developed the core Questaal theoretical framework.  J.B.\, J.M.\ performed Raman microscopy and photoluminescence measurements, contributed to single-crystal x-ray diffraction measurements, and led the analysis of the photoluminescence data. A.A.B. contributed to single-crystal measurements and structure modeling. J.J.\ contributed to experimental data acquisition, interpretation, figure preparation, and manuscript revision. All authors contributed to the analysis and critically revised the manuscript. 

\section*{Competing interests}
The authors declare no competing interests.

\section*{Ethics statement}
This study did not involve human participants, human data or tissue, or animal subjects, and therefore did not require ethics committee approval.

\begin{acknowledgments}
This work was authored by the National Laboratory of the Rockies for the U.S. Department of Energy (DOE) under Contract No.\ DE-AC36-08GO28308. S.A., J.M., J.C.J., and J.L.B. acknowledge funding from the Laboratory Discretionary Research and Development (LDRD) program of the National Laboratory of the Rockies.  For M.v.S. and D.P., funding was provided by the Computational Chemical Sciences program within the Office of Basic Energy Sciences, U.S.\ Department of Energy. S.A.\ acknowledges the use of computational resources sponsored by the Department of Energy's Office of Energy Efficiency and Renewable Energy, and located at the National Laboratory of the Rockies. S.A. acknowledges the use of the National Energy Research Scientific Computing Center, under Contract No. DE-AC02-05CH11231 using NERSC award BES-ERCAP0021783 for a portion of the work. The views expressed in the article do not necessarily represent the views of the DOE or the U.S.\ Government. The U.S.\ Government retains and the publisher, by accepting the article for publication, acknowledges that the U.S.\ Government retains a nonexclusive, paid-up, irrevocable, worldwide license to publish or reproduce the published form of this work, or allow others to do so, for U.S.\ Government purposes.
\end{acknowledgments}

% Bibliography for arXiv/REVTeX: ship manuscript.bbl (+ ref.bib). Do NOT \input the .bbl
% (REVTeX would load it twice and double the reference numbers).
\bibliographystyle{unsrt}
\bibliography{ref}

\clearpage
\appendix
\setcounter{figure}{0}
\setcounter{table}{0}
\setcounter{section}{0}
\setcounter{equation}{0}
\renewcommand{\thefigure}{S\arabic{figure}}
\renewcommand{\thetable}{S\arabic{table}}
\renewcommand{\thesection}{S\arabic{section}}
\renewcommand{\theequation}{S\arabic{equation}}
\captionsetup[figure]{labelformat=suppfig, labelsep=pipe}
\captionsetup[table]{labelformat=supptab, labelsep=pipe}

\section*{Supplementary Information}

\section{Crystal-growth images}

\begin{figure}[h!]
\centering
\includegraphics[width=0.78\linewidth]{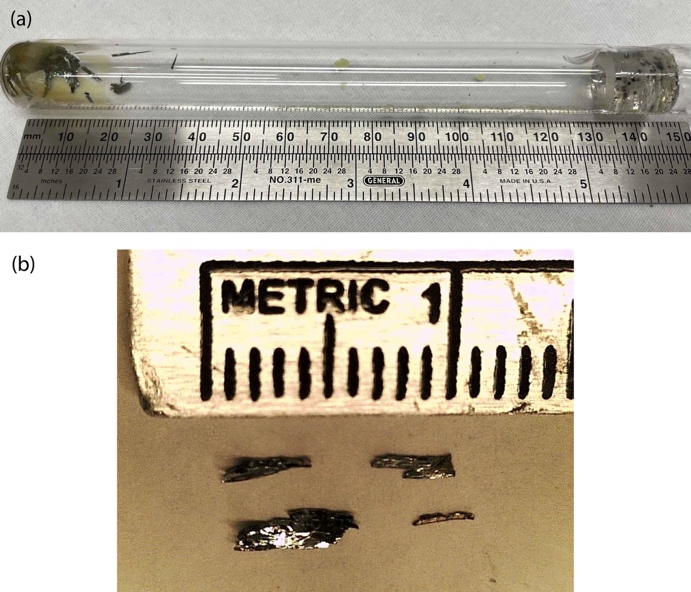}
\caption{Crystal-growth procedure and representative sample morphology. (a) Sealed quartz ampoule after the crystal-growth run. (b) Representative AgCrP$_2$S$_6$ crystals displaying the needle-like morphology used for single-crystal x-ray diffraction. A crop of panel (b) appears as Fig.~1(a) of the main text.}
\label{fig:si-synthesis}
\end{figure}

\section{Temperature-dependent Raman spectra}

Full-window Raman spectra at 108\,K, 140\,K, 175\,K, and 298\,K are collected here for completeness. The main text retains the 108\,K comparison together with the phonon dispersion and atom-projected phonon DOS.  Spectra at additional temperatures show that the same Raman windows persist across a range of temperatures.

\begin{figure*}[h!]
\centering
\includegraphics[width=0.8\linewidth]{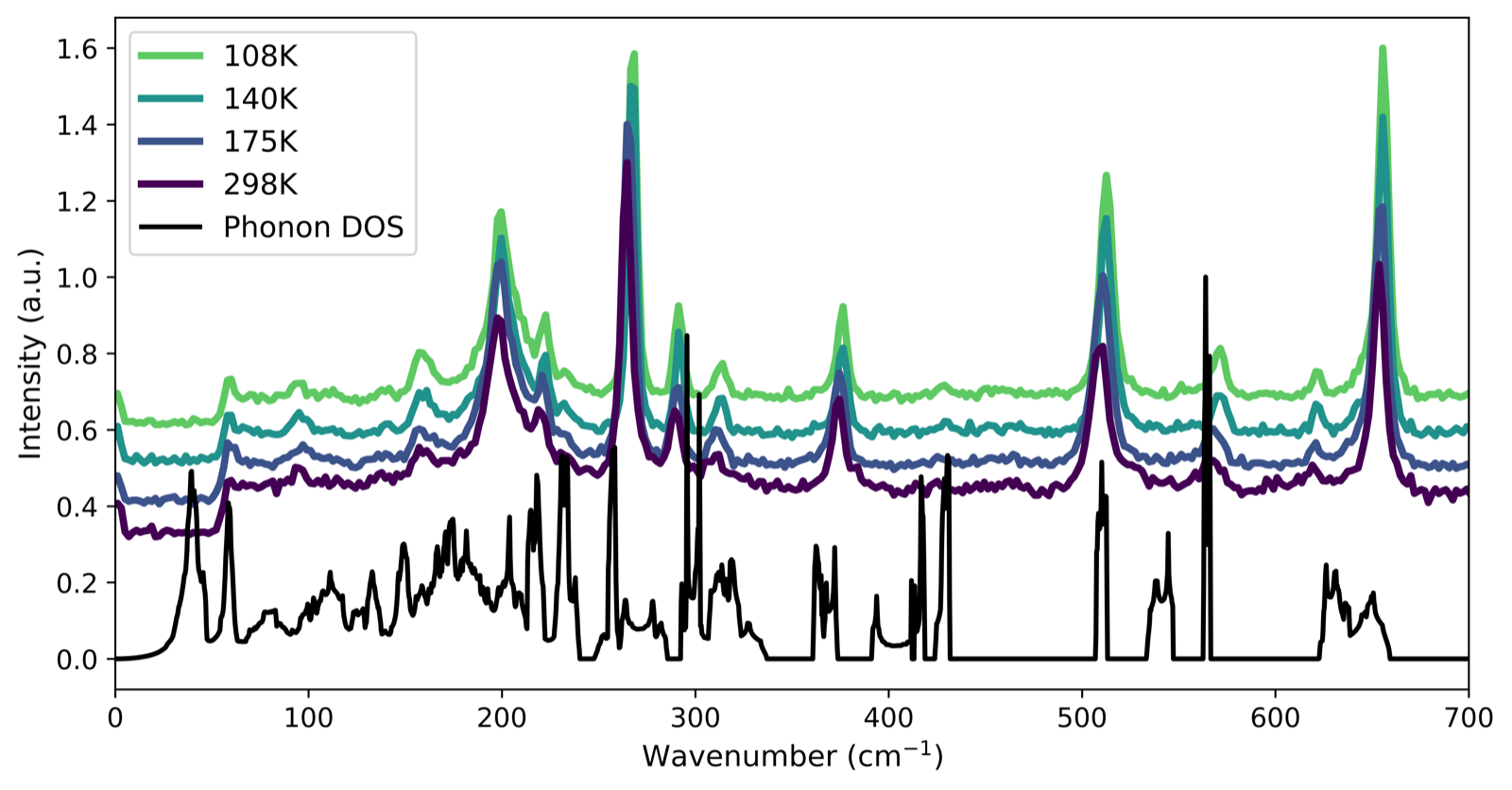}\hfill
\includegraphics[width=0.8\linewidth]{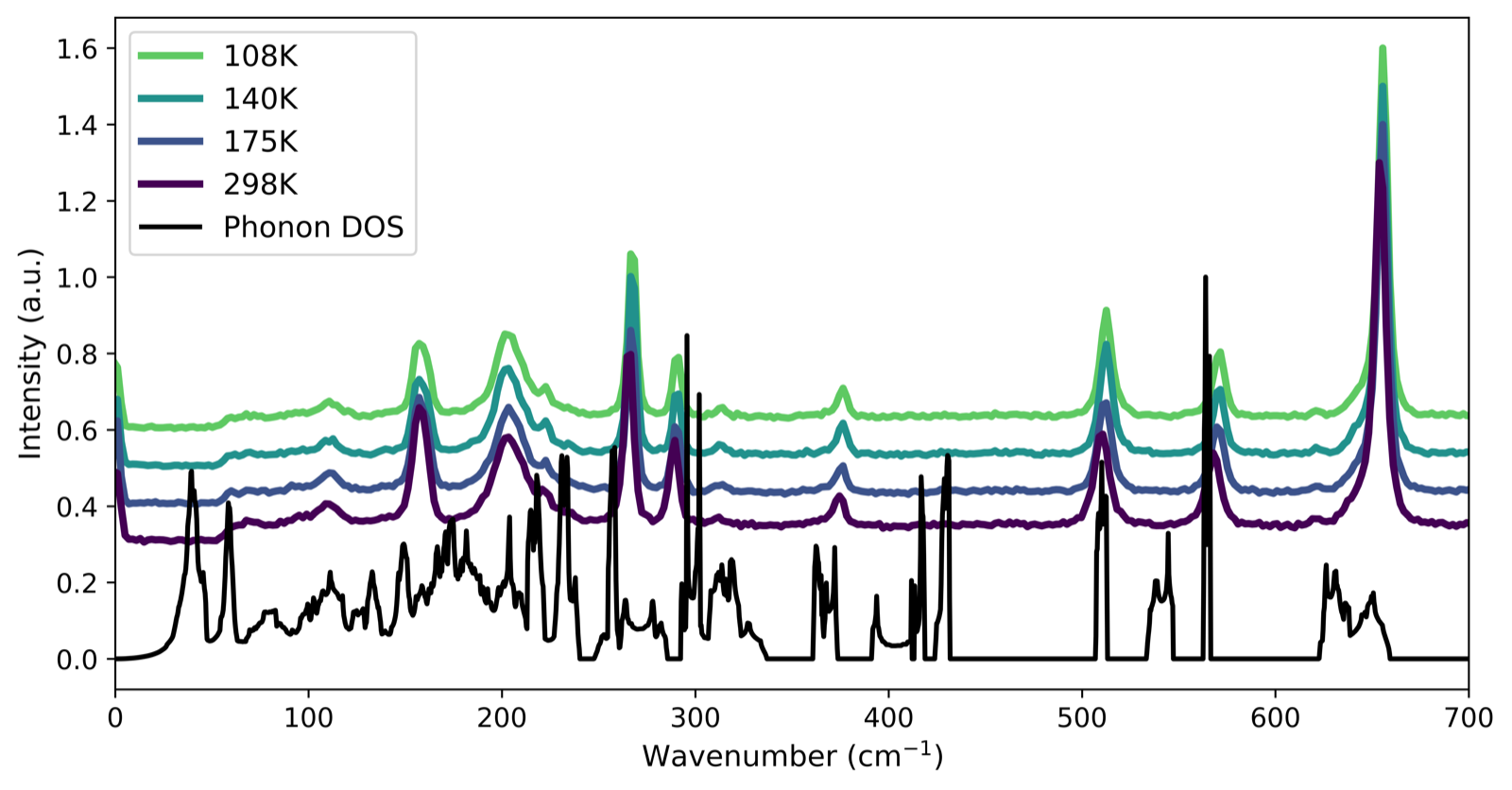}
\caption{Polarization-resolved Raman spectra of AgCrP$_2$S$_6$ compared with the calculated phonon density of states. Top: Polarized along the crystallographic \textit{a} axis. Bottom: Polarized along the crystallographic \textit{b} axis. Offsets are included for each measured spectrum.}
\label{fig:si-raman-full}
\end{figure*}

\section{Temperature-dependent photoluminescence spectra}

Photoluminescence was collected on a sample of AgCrP$_2$S$_6$ across a range of temperature values. The photoluminescence spectra were calibrated according to the procedure described in Appendix. 

\begin{figure*}[h!]
\centering
\includegraphics[width=0.49\linewidth]{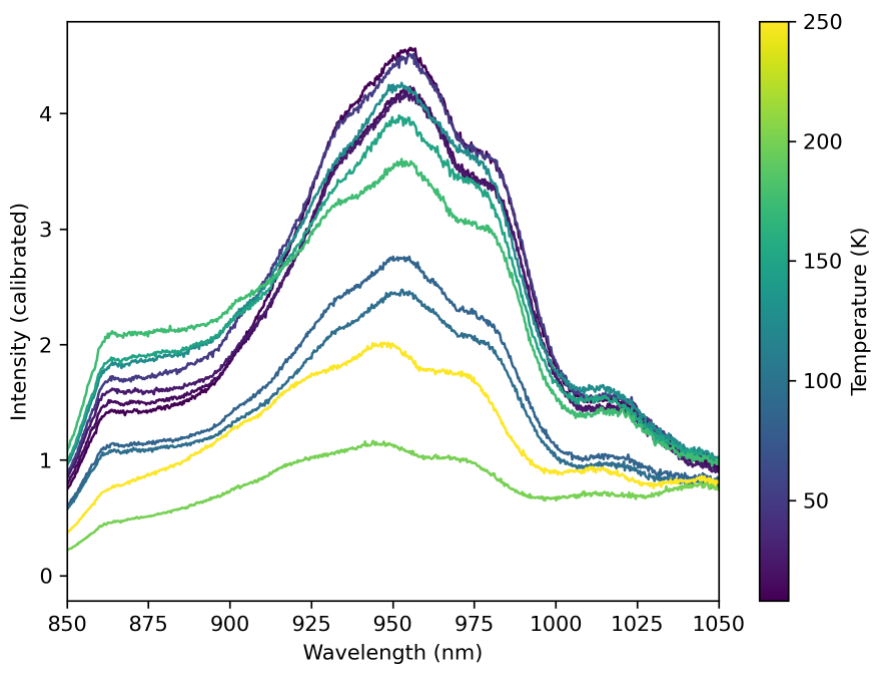}\hfill
\includegraphics[width=0.51\linewidth]{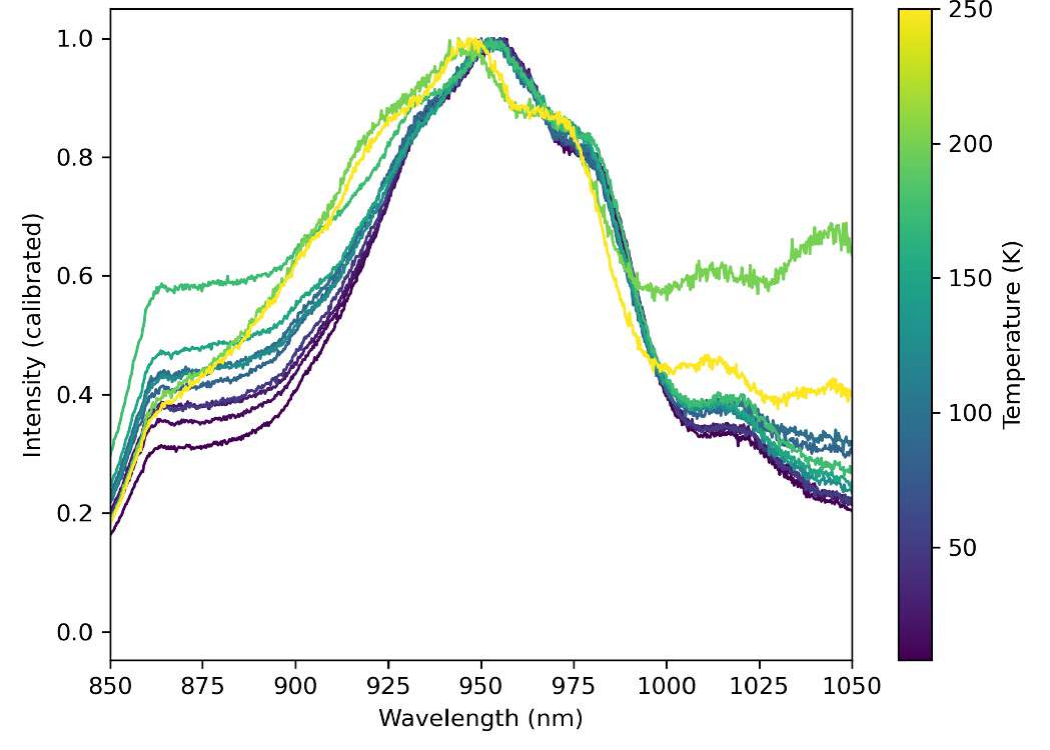}
\caption{Photoluminescence spectra of AgCrP$_2$S$_6$ across a range of temperatures. Left: Photoluminescence as collected, after calibration. Right: Photoluminescence as collected, after calibration, normalized by peak intensity.}
\label{fig:}
\end{figure*}

Photoluminescence data exhibit a moderate redshift with decreasing temperature, consistent with temperature-induced band gap renormalization. The structure of the photoluminescence peak is consistent across temperature. 

\section{Discussion of Photoluminescence Fitting Procedure and Exploration of Additional Models}

The Raman spectra of AgCrP$_2$S$_6$ (Figures 3c, S2) show several modes in the mid-frequency window relevant to the low-temperature photoluminescence. The PL envelope is structured by more than one phonon replica, so the vibronic progression was examined with single-mode and multi-mode Huang--Rhys fits as described below.

Symmetry analysis in the main text identifies exciton--phonon coupling channels for the lowest $X_A$ exciton. Combining that filter with our Raman spectra (Fig.~S2) and published low-temperature Raman frequencies leaves four candidate modes that are both Raman-active in our data and symmetry-allowed partners of $X_A$: $\nu_1$ = 267.5 cm$^{-1}$, $\nu_2$ = 291.4 cm$^{-1}$, $\nu_3$ = 313.6 cm$^{-1}$, and $\nu_4$ = 376.3 cm$^{-1}$. 

\begin{figure*}[h!]
\centering
\includegraphics[width=0.5\linewidth]{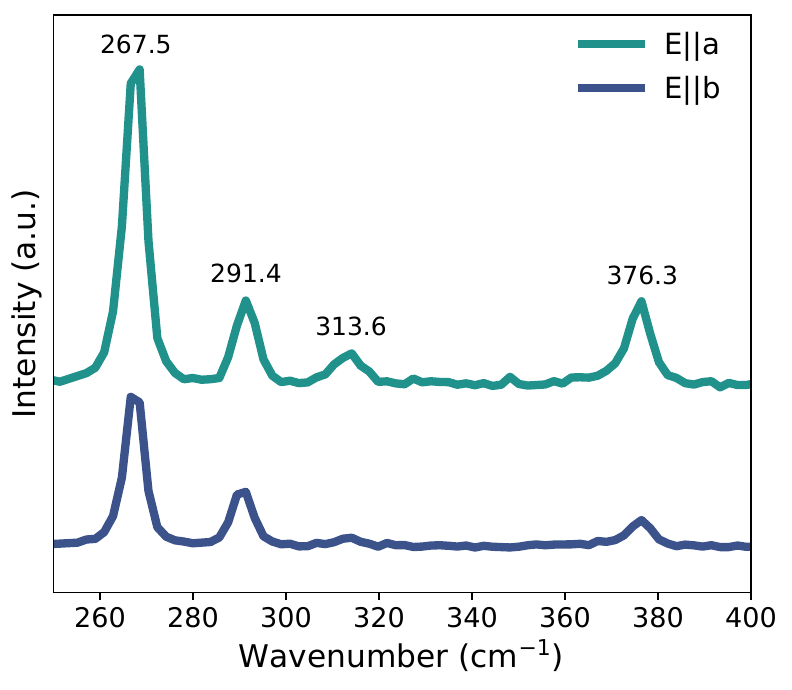}\hfill
\caption{Raman spectra at 108 K with Raman modes of interest labeled.}
\label{fig:si-raman-label}
\end{figure*}

Single-mode Huang--Rhys fits for each of these frequencies are shown in Figure~S5. They reproduce some of the sidebands but leave systematic residuals at other sub-peaks, so multi-mode progressions were considered next.

\begin{figure*}[h!]
\centering
\includegraphics[width=0.9\linewidth]{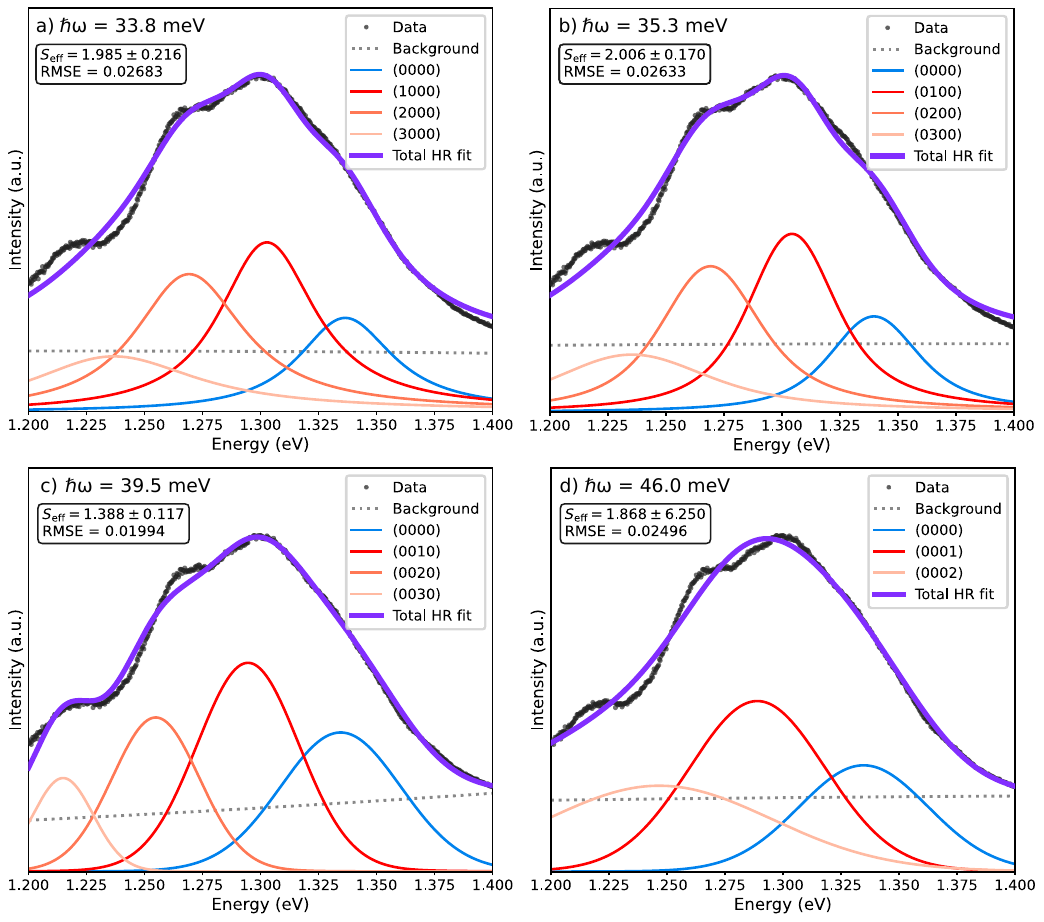}\hfill
\caption{Single-mode Huang-Rhys fits for the spectrum of interest, fitted to modes near the modes identified from Raman spectra (Figure S4).}
\label{fig:si-}
\end{figure*}

A multi-Voigt decomposition with independent intensities for the zero-phonon line and each sideband returns six peaks with the energies labeled in Figure~S7. Those spacings constrain which combinations of the four Raman modes can enter a Huang--Rhys progression. 

\begin{figure*}[h!]
\centering
\includegraphics[width=0.6\linewidth]{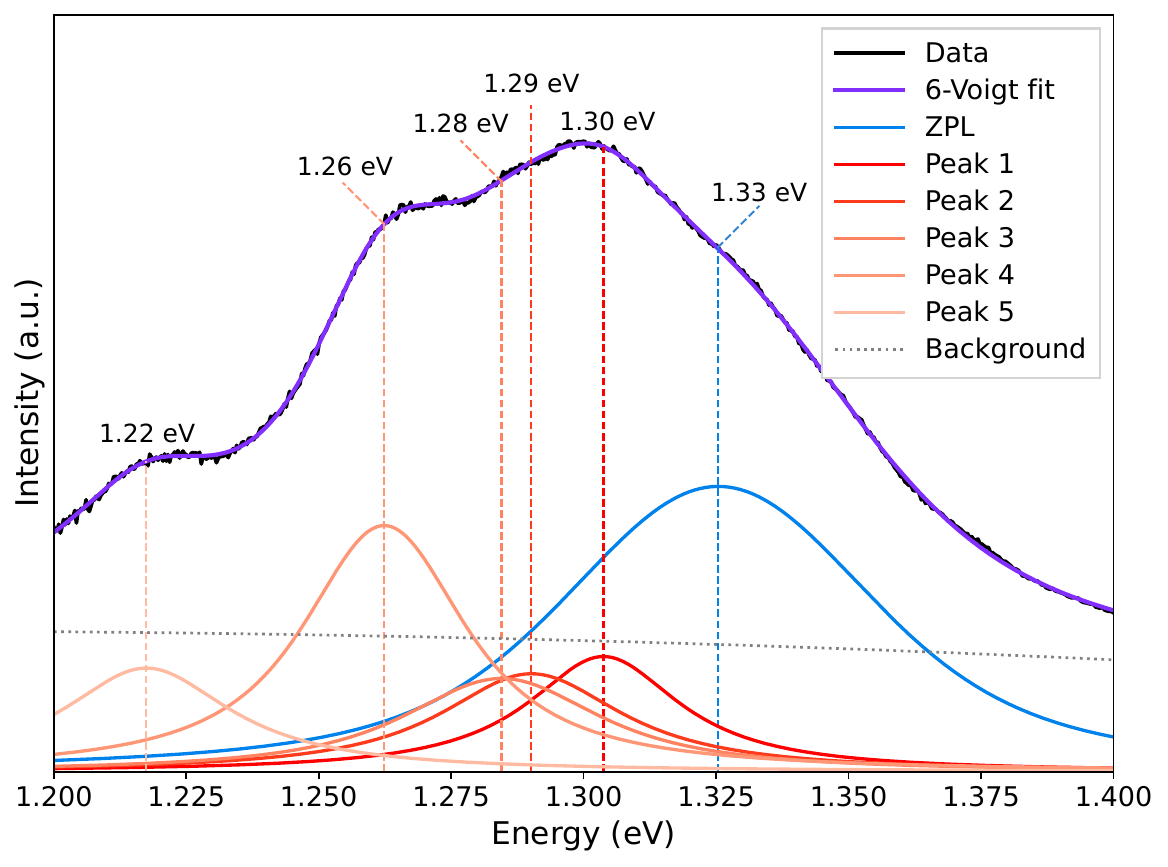}\hfill
\caption{Multi-Voigt fit of the photoluminescence spectrum, revealing sub-peak energies as labeled in the figure.}
\label{fig:si-raman-label}
\end{figure*}

Several Huang--Rhys progressions built from ($\nu_1$, $\nu_2$, $\nu_3$, $\nu_4$) match the Voigt spacings. Including the 291.4\,cm$^{-1}$ mode in some of the SI models changes the phonon indexing relative to the three-mode scheme in the main text. Progressions restricted to two or three distinct phonon modes were fitted; the combinations used here are:

\begin{enumerate}
\item Model A: (1000); (0010); (0001); (1010); (1011)
\item Model B: (1000); (0001); (1001); (2000); (2001)
\item Model C: (0100); (0001); (0200); (0101); (0002)
\item Model D: (0100); (0010); (0001); (0110); (0101); (0111)
\end{enumerate}

The least-squares fits for these models are shown in Figure S6.

\begin{figure*}[h!]
\centering
\includegraphics[width=0.9\linewidth]{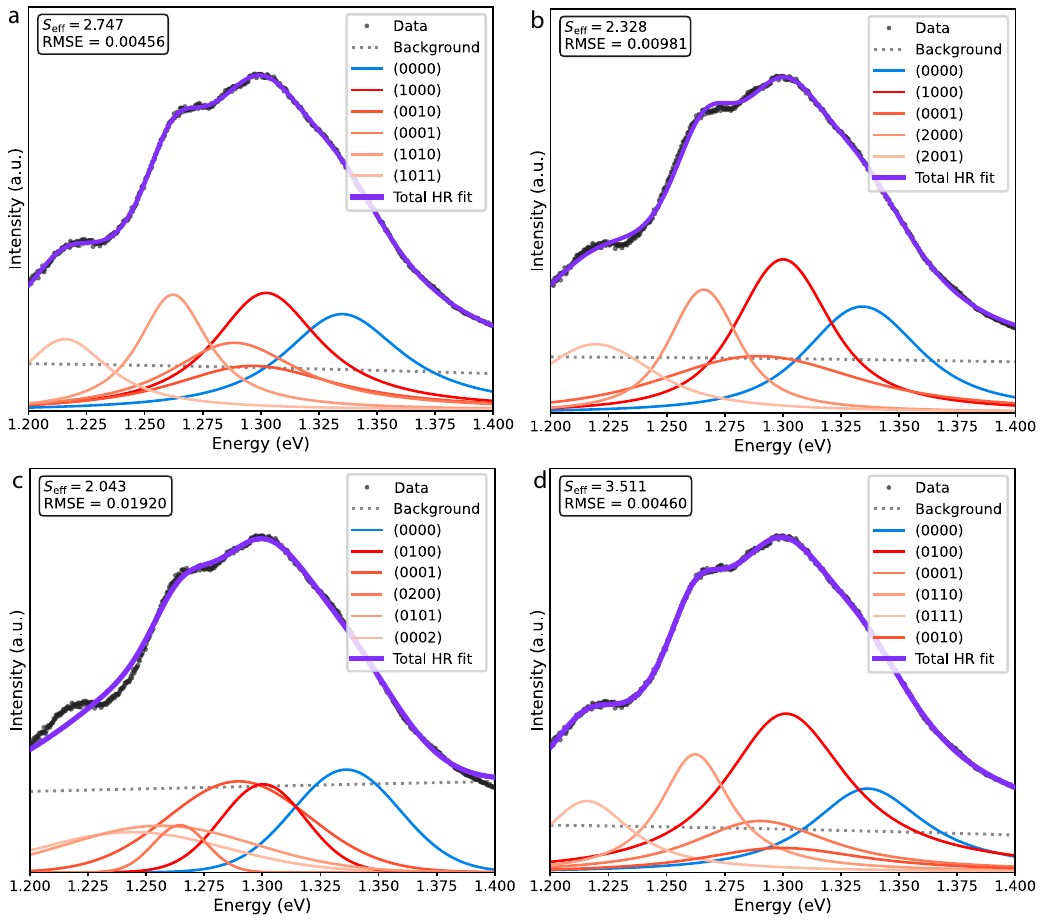}\hfill
\caption{Multi-mode Huang-Rhys fits for the spectrum of interest, fitted to various combinations of the modes of interest identified in the Raman spectra (Figure S4). Panels a, b, c, and d correspond to models A, B, C, and D, respectively. Details of the vibronic progression are shown in the legend for each model.}
\label{fig:si-raman-label}
\end{figure*}

The truncated models above omit some Franck--Condon combinations that fall inside the observed vibronic window. Expanded replica sets that include those additional combinations were therefore also fitted:

\begin{enumerate}
\item Model A Expanded: (1000); (0010); (0001); (2000); (1010); (0020); (1001); (0011); (0002); (3000); (2010); (1020); (0030); (2001)
\item Model B Expanded: (1000); (0001); (2000); (1001); (0002); (3000)
\item Model C Expanded: (0100); (0001); (0200); (0101); (0002); (0300); (0201)
\item Model D Expanded: (0100); (0010); (0001); (0200); (0110); (0020); (0101); (0011); (0002); (0300); (0210); (0120); (0030); (0201); (0111)
\end{enumerate}

\begin{figure*}[h!]
\centering
\includegraphics[width=0.9\linewidth]{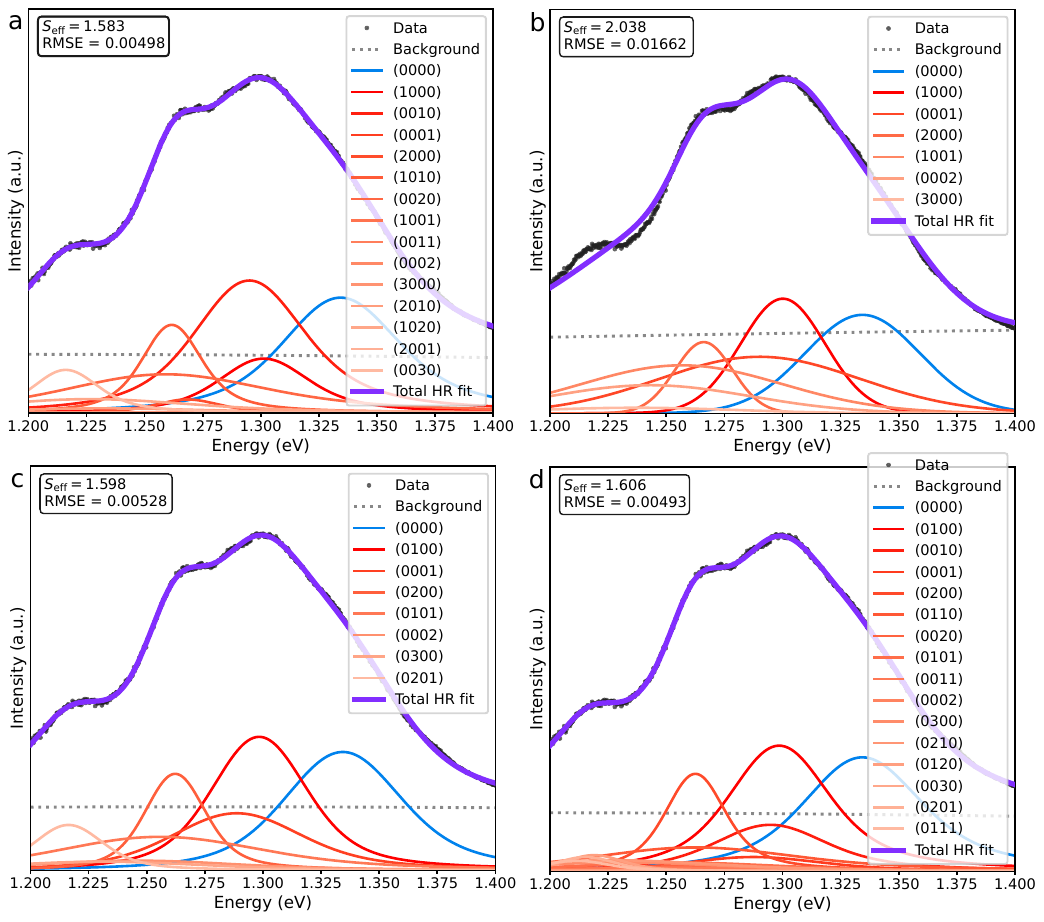}\hfill
\caption{Multi-mode Huang-Rhys fits for the spectrum of interest, fitted to various combinations of the modes of interest identified in the Raman spectra (Figure S4). Panels a, b, c, and d correspond to expanded models A, B, C, and D, respectively. Spectra display a truncated selection of all of the energy modes expected to be present in this spectrum according to a Franck-Condon picture. Details of the vibronic progression are shown in the legend for each model.}
\label{fig:si-model-expand}
\end{figure*}

\begin{table}[]
\begin{tabular}{|c|c|c|c|c|}
\hline
\textbf{Model} & \textbf{\(S_{\mathrm{eff}}\)} & \textbf{RMSE} & \textbf{AIC} & \textbf{BIC} \\ \hline
Model A & 2.764 & 0.00456 & $-$6074.83 & $-$5992.36 \\
Model A Expanded & 1.583 & 0.00498 & $-$5957.37 & $-$5835.84 \\
Model B & 1.825 & 0.01028 & $-$5157.21 & $-$5083.42 \\
Model B Expanded & 2.038 & 0.01662 & $-$4609.87 & $-$4531.74 \\
Model C & 2.043 & 0.01920 & $-$4448.73 & $-$4374.94 \\
Model C Expanded & 1.598 & 0.00528 & $-$5908.58 & $-$5826.12 \\
Model D & 4.862 & 0.00462 & $-$6057.35 & $-$5970.54 \\
Model D Expanded & 1.606 & 0.00493 & $-$5967.40 & $-$5841.53 \\ \hline
\end{tabular}
\caption{Effective Huang-Rhys factors, root mean square error (RMSE), Akaike Information Criterion (AIC), and Bayesian Information Criterion (BIC) resulting from the least-squares regression of the photoluminescence data to the eight models of interest. Model A yields the lowest values of all three fit statistics among the models compared.}
\end{table}

The expanded fits are shown in Fig.~S\ref{fig:si-model-expand}. Fit statistics for each model and its expanded counterpart are collected in Table~S1.

Among the eight fits, the non-expanded Model~A (zero-phonon line plus five vibronic subpeaks) gives the lowest RMSE (0.00456) and the lowest AIC and BIC, and is the progression reported in the main text. Across the set, the effective Huang--Rhys factor $S_{\mathrm{eff}}$ spans roughly 1.6 to 4.9. In every case $S_{\mathrm{eff}}$ remains of order unity or larger. Figure~S9 shows 95\% confidence intervals for $S_{\mathrm{eff}}$ obtained from the rise in the residual sum of squares relative to its unconstrained minimum; for most models those intervals include values between 2 and 2.5.

\begin{figure*}[h!]
\centering
\includegraphics[width=0.7\linewidth]{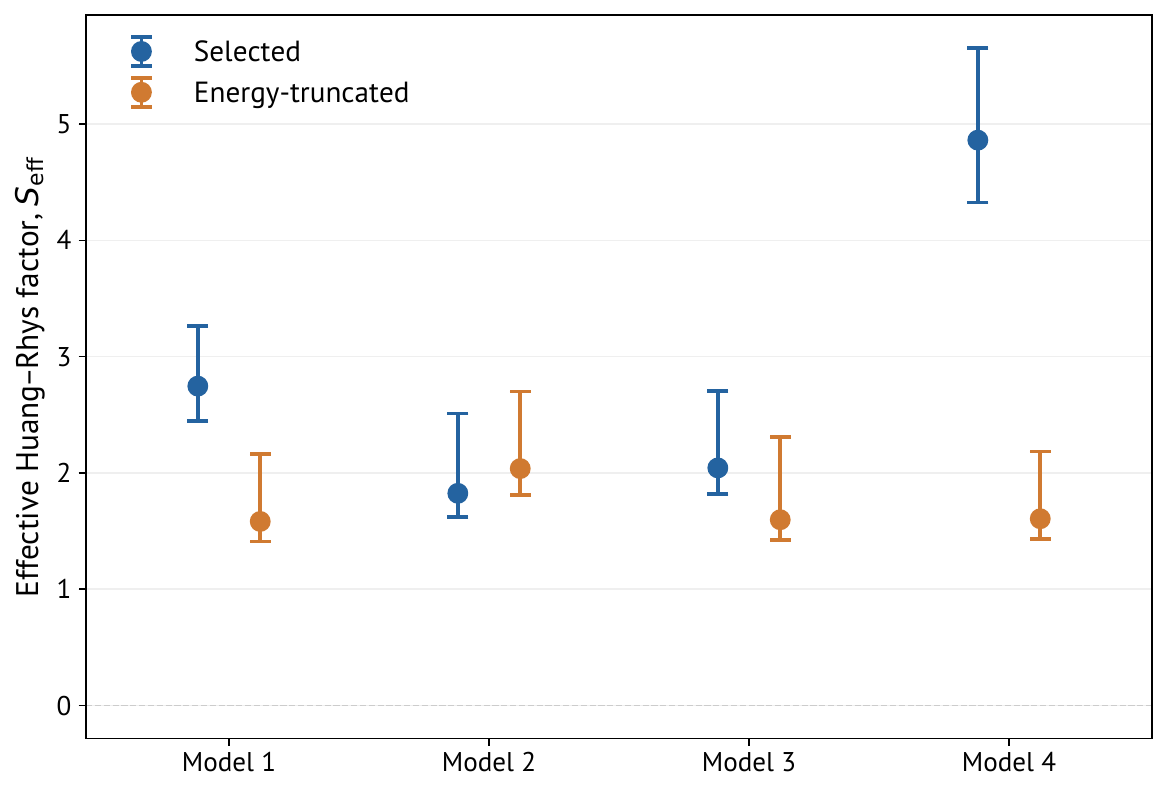}\hfill
\caption{Effective Huang-Rhys factor for each model of interest, with a 95\% confidence interval as extracted from the fit.}
\label{fig:si-raman-label}
\end{figure*}

Expanding the replica set lowers $S_{\mathrm{eff}}$ for three of the four base models. Some expanded replicas return near-zero fitted intensity. Neighboring replicas overlap strongly in energy, so those amplitudes are correlated in the least-squares fit and individual Huang--Rhys weights for crowded replicas are less well determined than the overall $S_{\mathrm{eff}}$.

Because several Raman-active modes lie close in energy within the $X_A$ coupling window, more than one vibronic progression can describe the PL envelope. Across the models examined here, $S_{\mathrm{eff}}$ stays in the moderate-to-strong range, as expected for a localized lowest exciton.

\end{document}